\documentclass{achemso}
\setkeys{acs}{usetitle=true}

\usepackage[version=3]{mhchem} 
\usepackage{multirow}

\author{Trang N. Do}
\affiliation{SISSA/ISAS - International School for Advanced Studies, Trieste 34136, Italy}
\author{Paolo Carloni}
\affiliation{Computational Biophysics, German Research School for Simulation Sciences, D-52425 J\"ulich, Germany and Institute for Advanced Simulation IAS-5, Computational Biomedicine, Forschungszentrum J\"ulich, D-52425 J\"ulich, Germany}
\author{Gabriele Varani}
\affiliation{Department of Chemistry and Department of Biochemistry, University of Washington, Seattle, Washington 98195-1700, United States}
\author{Giovanni Bussi}
\email{bussi@sissa.it}
\affiliation{SISSA/ISAS - International School for Advanced Studies, Trieste 34136, Italy}

\title{RNA/peptide binding driven by electrostatics $-$ Insight from bi-directional
pulling simulations}

\begin{document}

\begin{abstract}
RNA/protein interactions play crucial roles in controlling gene expression.
They are becoming important targets for pharmaceutical applications.
Due to RNA flexibility and to the strength of electrostatic interactions,
standard docking methods are insufficient. We here present a computational
method which allows studying the binding of RNA molecules and charged
peptides with atomistic, explicit-solvent molecular dynamics. In our
method, a suitable estimate of the electrostatic interaction is used
as an order parameter (collective variable) which is then accelerated
using bi-directional pulling simulations. Since the electrostatic
interaction is only used to enhance the sampling, the approximations
used to compute it do not affect the final accuracy. The method is
employed to characterize the binding of TAR RNA from HIV-1 and a small
cyclic peptide. Our simulation protocol allows blindly predicting
the binding pocket and pose as well as the binding affinity. The method
is general and could be applied to study other electrostatics-driven
binding events.
\end{abstract}

\section{Introduction}

Peptidic ligands are providing increasingly attractive drug leads
for many proteins. They have recently been applied to RNA as well \cite{FroeyenHerdewijn2002,Zacharias2003,Vicens2009}.
Targeting RNA is, however, more challenging than targeting proteins
due to the highly charged nature and flexibility of RNA \cite{Puglisi1993,AuffingerWesthof2000,Hermann2002,HermannTor2005,AuffingerHashem2007}.
In addition, designing RNA-binding drugs is limited by the poor understanding
of RNA/ligand molecular recognition events \cite{GallegoVarani2001,Bosshard2001,Boehr2009,FulleGohlke2010}.
The latter are intricate processes strongly
dominated by a conformational selection and/or induced fit mechanisms \cite{LeulliotVarani2001,AlHashimi2005}.
Structural information by X-ray and NMR experiments is crucial to
provide insights on the binding poses of RNA/peptide complexes \cite{Hamy1997}.
NMR techniques further give useful information on dynamics \cite{LipariSzabo1982-1,LipariSzabo1982-2}. 
Computational approaches have proved to be useful for studying nucleic-acid/ligand
complexes. On one hand, docking approach using coarse-grained modeling
allows efficient and systematic search for the native-like binding
poses \cite{Setny2012}. On the other hand, Molecular Dynamics (MD)
simulations with implicit-solvent representations have shown to provide
a relatively good agreement with experiments in structural and dynamical
properties of the complexes \cite{BanavaliRoux2002,Rizzo2006,FulleGohlke2010}.
Finally, using an explicit representation of solvent and ions could
shed light on the RNA structural adaptation and molecular recognition
under realistic conditions of the ionic solution, which is known to
play important roles in RNA stability and inter-molecular electrostatic
interactions \cite{Draper2005,Chen2009}.

However, an explicit representation
of solvent and ions implies a significant increment in the number
of atoms to be simulated \cite{MackerellNilsson2008}. Therefore,
a well-known difficulty of atomistic MD simulations is that they can
only follow the system dynamics on the microsecond timescale at most.
Although recent developments allowed reaching the millisecond timescale
at least for small proteins \cite{Shaw2010,Lindorff2011}, the study
of slower conformational transitions of larger molecules require some
form of acceleration. Several methods have been proposed to alleviate this problem, and many of them are based on the
\emph{a priori} choice of proper reaction coordinates, or Collective Variables (CVs), which are biased
during the simulation (see e.g. \cite{TorrieValleau1977,Mezei1987,Grubmuller1996,DarvePohorille2001,LaioParrinello2002,Rosso2002,Marsili2006,Maragliano2006,SotomayorSchulten2007,Barducci2008}.)
A common difficulty of these algorithms is to properly choose an effective CV which is highly
dependent on the specific problem. Several authors have proposed to
bias the potential energy of the system as a solution \cite{BartelsKarplus1998,WangLandau2001,Micheletti2004,Michel2009,BonomiParrinello2010}.
However, the potential energy should not be interpreted as a proper
CV in solvated systems where large fluctuating contributions arise
from the solvent-solvent interactions. Indeed, these methods are efficiently
used in solvated systems in a spirit more related to that of parallel
tempering \cite{SugitaOkamoto1999}, simulated tempering \cite{MarinariParisi1992},
and multicanonical sampling \cite{BergNeuhaus1992}, i.e. to let the
system evolve in an ensemble where effective barriers are decreased
and conformational transitions are more likely to occur. In this respect,
it appears fascinating to push this idea further and to use as a CV
only the component of the energy which is relevant for the transition
of interest such that the solvent fluctuations are averaged out.

In this paper, we propose to use an approximation of the electrostatic
interaction free-energy between two molecules as a CV for binding
problems. This free energy can be easily computed on the fly within
the Debye-H\"uckel formalism and can be used as a descriptor to distinguish
the bound (low free energy) and unbound (high free energy) states.
The so-called Debye-H\"uckel energy (DHEN) is only an approximate evaluation
of the intermolecular interactions, and is here merely used as a CV
on top of which a suitable bias is added. In this manner, the accuracy
of the free energy calculation is not dictated
by the Debye-H\"uckel approximation.

\begin{figure}
\includegraphics[width=0.6\textwidth]{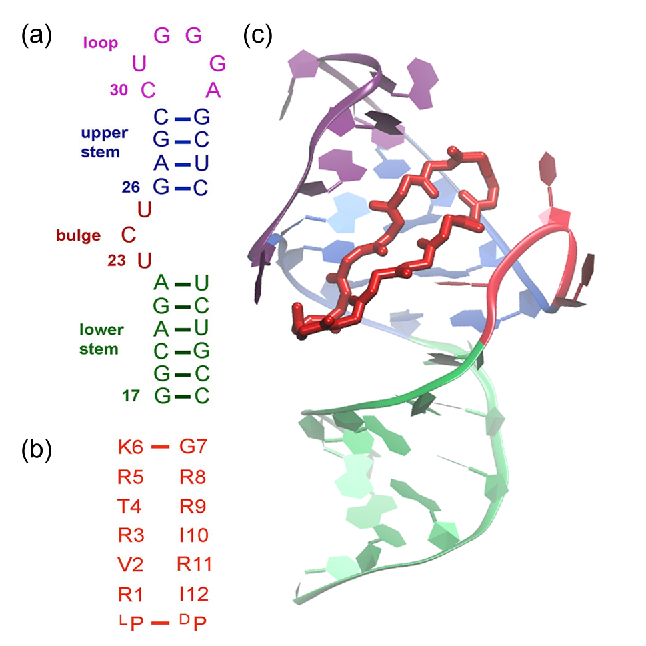}
\caption{(a) Apical portion (nucleotides G17 \textendash{} G45) of HIV-1 TAR
RNA; (b) Sequence of the so-called L22 peptide, a cyclic peptide mimic
of Tat protein; (c) NMR structure of the TAR/L22 complex.}
\label{Structures}
\end{figure}

We here perform bi-directional Steered Molecular Dynamics (SMD) simulations
\cite{Grubmuller1996,SotomayorSchulten2007} with bias applied on
our proposed DHEN CV. The case-study system is the HIV Trans-Activation
Responsive (TAR) RNA element in complex with a cyclic peptide inhibitor.
TAR is a stem-bulge-loop RNA structure, formed by the first 59 nucleotides
of the nascent HIV-1 mRNA molecules. Its apical portion (nucleotides
G17 - G45, see Figure \ref{Structures}(a) directly binds to the arginine-rich
viral trans-activator of transcription protein (Tat), enhancing a
key step in the viral replication, i.e. the transcription from the
pro-viral DNA into the full-length viral mRNA \cite{Dingwall1989,Dingwall1990,Weeks1990,Calnan1991}.
Therefore, the TAR/Tat interaction is a potential target for developing
new anti-HIV drugs. Due to conserved structure of TAR \cite{Delling1992,Huthoff2001},
drug resistance development against TAR-binding inhibitors is expected
to be slower than with other conventional HIV-1 protein targets. As
a new approach to tackling TAR/Tat interaction, Davidson et al., focused
on the development of conformationally constrained mimics of HIV-1
Tat and discovered a family of beta-hairpin cyclic peptides that bind
to TAR with nanomolar affinity and greatly improved specificity compared
with previous ligands \cite{Davidson2009}. Among more than 100 peptides
in the investigated family, the arginine-rich sequence cyclo-RVRTRKGRRIRIPP
(L22 hereafter, see Figure \ref{Structures}(b)) stands out for its
potency. L22 binds to TAR with the affinity of 1 nM. NMR experiments
also showed that L22 peptide binds to the major groove of the upper
RNA helix (nucleotides 26-29 and 36-39, see Figure \ref{Structures}(c)),
which is also the binding pocket of Tat \cite{Dingwall1989,Weeks1990}.

In a recent work, we employed atomistic MD simulations to investigate
structural and dynamical properties of the TAR/L22 complex \cite{Do2012}.
Our simulations also confirmed that binding between an RNA and a positively-charged
peptide is a spontaneous process strongly driven by electrostatic
interactions. In this work, using bi-directional SMD simulations with
our proposed DHEN CV, we are able to predict the correct binding pocket
and binding pose without any \textit{a priori} structural information
of the complex. We here also generalize a reweighting technique to
compute the Potential of Mean Force (PMF) in a bi-directional steering
scheme on any \textit{a posteriori} chosen CV, which is not necessarily
the steered CV.


\section{Materials and Methods}

\subsection{Theoretical Approaches}

\subsubsection{Electrostatic Interaction as a Collective Variable}
We construct a CV describing in an approximate manner the electrostatic
interaction free energy between the ligand $B$ and its target $A$.
Our CV aims to be highly general, computationally efficient, and to
use a small number of parameters. This obviously comes at the cost
of accuracy, which appears to be well justified here because the approximate
electrostatic free energy of the system is used only as a CV for guiding
the exploration of the conformational space.

To derive our CV, we make several simplifying assumptions: \textit{(i)}
the ionic solution is highly dilute such that the relation $e_{c}\Phi(\mathbf{r})\ll k_{B}T$
holds, here $e_{c}$, $k_{B}$, $T$ are the constants denoting the
electronic charge, Boltzmann constant, and temperature of the system
respectively, and $\Phi(\mathbf{r})$ represents the electrostatic
potential at a position $\mathbf{r}$; \textit{(ii)} the difference
in electrostatic interactions due to different atomic sizes are negligible;
and \textit{(iii)} the ionic solution is considered homogeneous; i.e.
the decreasing in solution-screening effect at the region close to
or inside the molecules, which causes an increasing in strength of
electrostatic interactions, is also negligible. These assumptions do not affect the accuracy of free-energy calculations, which is obtained from the work performed along steered MD simulations.

Let us consider the electrostatic potential $\Phi(\mathbf{r})$ of
a target/ligand complex system in an aqueous solution containing a
dilute 1:1 electrolyte (K$^{+}$ and Cl$^{-}$ featuring an ionic
strength of 10 mM in our case). The electrostatic potential can be
calculated by the linearized Poisson-Boltzmann equation (PBE) or the
Debye-H\"{u}ckel equation \cite{Leach2001}:

\begin{equation}
-\nabla(\epsilon(\mathbf{r})\nabla\Phi(\mathbf{r}))=4\pi\sum_{i=1}^{N_{m}}q_{i}\delta(\mathbf{r}-\mathbf{r}_{i})-\bar{\kappa}^{2}(\mathbf{r})\Phi(\mathbf{r}),\label{linearizedPBE}
\end{equation}
where the permittivity $\epsilon(\mathbf{r})$ adopts appropriate
dielectric constant values in different regions of the Debye-H\"uckel
model; the molecule is composed of $N_{m}$ atoms $i$ at positions
$\mathbf{r}_{i}$ with point charges $q_{i}$ represented by the Dirac
delta functions; and $\bar{\kappa}(\mathbf{r})$ is the modified dielectric-independent
Debye-H\"uckel parameter defined as $\bar{\kappa}(\mathbf{r})=\sqrt{\epsilon_{w}}\kappa$
if $\mathbf{r}$ belongs to the solvent region of Debye-H\"uckel model
and $\bar{\kappa}(\mathbf{r})=0$ elsewhere, $\epsilon_{w}$ is the
dielectric constant of water and $\kappa$ is the usual Debye-H\"uckel
parameter whose value is given by:

\[
\kappa=\Bigl(\dfrac{8\pi N_{A}e_{c}^{2}}{1000\epsilon_{w}k_{B}T}\Bigl)^{1/2}I_{s}^{1/2},
\]
where $N_{A}$ is the Avogadro number and $I_{s}=1/2\sum_{i=1}^{N_{I}}c_{i}z_{i}^{2}$
is the ionic strength of the solution determined by $N_{I}$ types
of ions, each type has a charge $q_{i}=z_{i}e_{c}$ and a concentration
$c_{i}$.

The general analytical solution of Equation (1)
in the solvent region is given by
\begin{equation}
\Phi(\mathbf{r})=\frac{1}{k_{B}T\epsilon_{w}}\sum_{i=1}^{N_{m}}\frac{q_{i}e^{-\kappa|\mathbf{r}-\mathbf{r}_{i}|}}{|\mathbf{r}-\mathbf{r}_{i}|},\label{potential}
\end{equation}
From the electrostatic potential in Equation (2), one
can easily derive the electrostatic-interaction term in the free energy
of a two-non-overlapping-molecule system as:
\begin{equation}
G^{DH}=\sum_{j\in B}q_{j}\Phi(\mathbf{r}_{j})=\frac{1}{k_{B}T\epsilon_{w}}\sum_{j\in B}\sum_{i\in A}q_{i}q_{j}\frac{e^{-\kappa|\mathbf{r}_{ij}|}}{|\mathbf{r}_{ij}|},\label{DHEN}
\end{equation}
where A (B) is the set of all the atoms of the target (ligand) molecule;
$i$ and $j$ are the atom indexes in the two sets $A$ and $B$;
$|\mathbf{r}_{ij}|=|\mathbf{r}_{i}-\mathbf{r}_{j}|$ denotes the distance
between atoms $i$ and $j$. Here we introduce a further simplification
by assuming that only the inter-molecular electrostatic interactions
contribute significantly to the free-energy difference between the
bound and unbound states of a complex. We thus ignore the intramolecular
relaxation in the Equation (3).

\subsubsection{Validation against Poisson-Boltzmann Equation}
To compare the estimated electrostatic free energy using nonlinear
and linearized PBEs, we performed electrostatic calculations on our
case study, the L22-TAR complex system. To build our initial molecular
configuration, we first placed L22 at the origin of an arbitrarily
selected coordinate system and then placed TAR at 40 Å on an axis,
e.g. the $x$-axis in our setup. This choice of distance ensured that
L22 and TAR were far enough to have no inter-molecular contacts. The
dipole moments of both molecules were aligned with another direction,
e.g. the $z$-axis in our setup. We next progressively rotated both
L22 and TAR molecules about the $x$-, $y$-, and $z$- axes. The
combination of these rotations is equivalent to placing the ligand
in all three-dimensional rotations everywhere on the surface of a
sphere with a radius of 40 Å around the RNA. The ``angle-step''
of the rotation was $1^{\circ}$. At every step, the electrostatic-interaction
free energy was calculated by two methods: \textit{(i)} numerically
solving the nonlinear PBE using APBS 1.3 \cite{APBS2001} and \textit{(ii)}
using Equation (3), which is resulted from the analytical
solution of the linearized PBE. These calculations provided the orientation
dependence of the electrostatic interaction.

\begin{figure}
\includegraphics[width=0.6\textwidth]{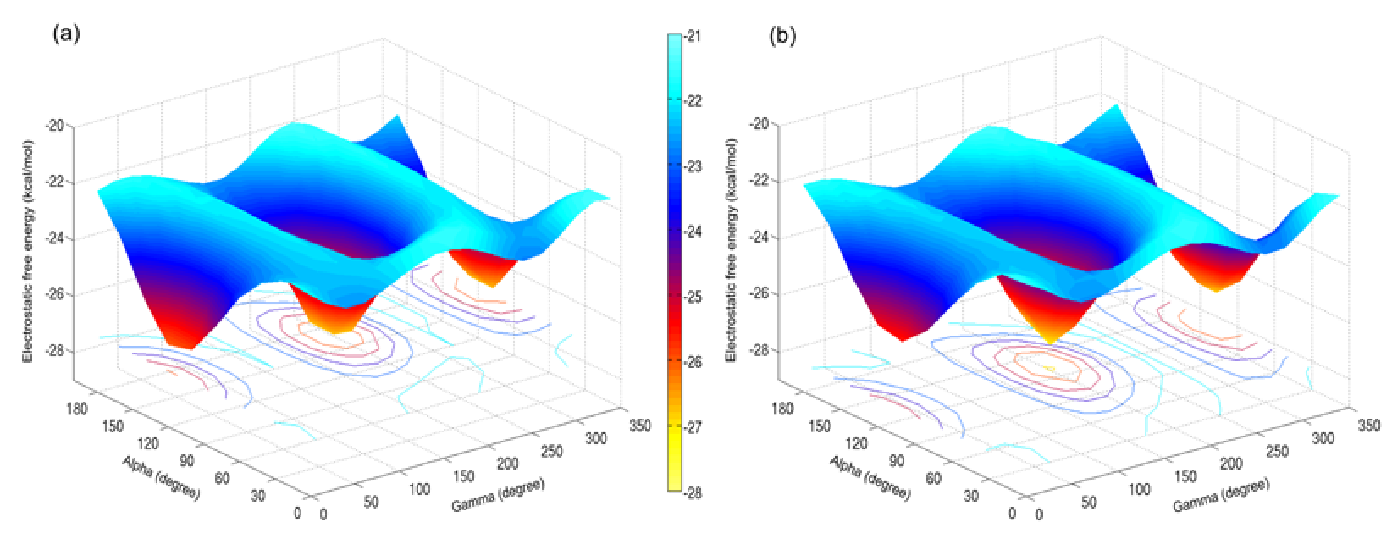}
\caption{Electrostatic interaction free energy as a function of TAR's orientations
calculated by both methods: solving nonlinear PBE (a) and linearized
PBE (b). Both calculations agree on two energy minima corresponding
to two orientations of TAR at which L22 face both the upper and lower
major groove of TAR. }
\label{APBSvsDHEN}
\end{figure}

\begin{figure}
\includegraphics[width=0.6\textwidth]{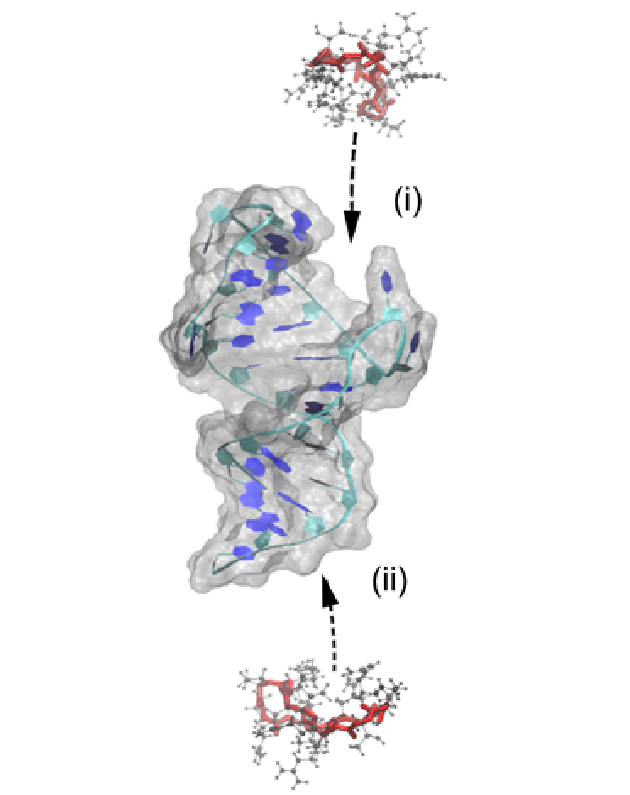}
\caption{Possible approaches of L22 to TAR predicted by electrostatic-free-energy
calculations: (i) upper major groove, which is as well the binding
site of Tat and (ii) lower major groove.}
\label{Approaching}
\end{figure}

Obviously, the performance of numerical PBE solving depends on the softwares,
grid spacing, and other procedures as well as input parameters.
Although we did not optimize our PBE-solving procedures,
the evaluation of our DHEN CV has a clear computational advantage compared to
solving PBE numerically. The results, however,
were not considerably different from one another. That can be observed
in Figure \ref{APBSvsDHEN}, which shows the dependence of electrostatic
free energy on the two angles $\alpha$ and $\gamma$ describing the
rotation of TAR about the $x$- and $y$- axes. Both calculation methods
agreed on the two minima representing two orientations of TAR at which
the electrostatic free energy of the system had the lowest values.
In these configurations, L22 faced both the upper and lower major
groove of TAR. This result further suggested that there are two possible
low-energy funnels in which L22 can approach TAR (see Figure \ref{Approaching}).
This agreement can become weaker as the two molecules
get closer and their low dielectric interiors approach each other.
However, under this condition, the full solution of the nonlinear
PBE could underestimate the change in molecular flexibility,
thus giving a poor estimate of the interaction free energy anyway.
We therefore believe that our DHEN provides a good compromise between
accuracy and computational cost.
These observations provided us with more confidence when using
the electrostatic free energy in Equation (3) as a CV to
accelerate the simulations.

We therefore propose to use the expression (3), hereafter
referred to as Debye-H\"uckel free energy (DHEN), as a CV. The DHEN
CV was implemented in an in-house version of PLUMED 1.3 \cite{PLUMED2009}.

\subsubsection{Reconstruction of the PMF from Bi-directional Pullings}
\label{FESReconstruction}

\paragraph{PMF as a Function of the Steered CV}
To reconstruct the PMF as a function of
the steered CV, we employed the bi-directional PMF estimator developed
by Minh and Adib \cite{MinhAdib2008}:
\begin{equation}
G_{0}(z)=-\beta^{-1}\ln\frac{\sum_{t}\left[\left\langle \frac{n_{F}\delta(z-z_{t})e^{-\beta W_{0}^{t}}}{n_{F}+n_{B}e^{-\beta(W-\triangle F)}}\right\rangle _{F}+\left\langle \frac{n_{B}\delta(z-z_{\tau-t})e^{\beta W_{\tau-t}^{\tau}}}{n_{F}+n_{B}e^{\beta(W+\triangle F)}}\right\rangle _{B}\right]e^{\beta\triangle F_{t}}}{\sum_{t}e^{-\beta\left[V(z;t)-\triangle F_{t}\right]}},\label{MinhAdib}
\end{equation}
where $G_{0}(z)$ denotes the PMF as a function of the CV $z$; $\tau$
is the total time of each pulling; $z_{t}$ is the value of the CV
$z$ at time $t$; $\langle\;\;\rangle_{F}$ and $\langle\;\;\rangle_{B}$
denote the averages taken over all forward and backward realizations
respectively; $n_{F}$ and $n_{B}$ are the number of realizations
in forward and backward pullings; $W_{0}^{t}$ is the cumulative pulling
work at time $t$; $W$ is the total work at time $\tau$; $V(z;t)=k[z-z_{\textrm{restr}}(t)]^{2}/2$
is the harmonic potential acting on the CV at time $t$, where $z_{\textrm{restr}}(t)$
denotes the value of the restrained CV; $\triangle F_{t}$ is the
free-energy difference between the equilibrium state at time $t$
and the initial equilibrium state of the forward process, whose value
is given by:

\begin{equation}
e^{-\beta\triangle F_{t}}=\left\langle \frac{n_{F}e^{-\beta W_{0}^{t}}}{n_{F}+n_{B}e^{-\beta(W-\triangle F)}}\right\rangle _{F}+\left\langle \frac{n_{B}e^{\beta W_{\tau-t}^{\tau}}}{n_{F}+n_{B}e^{\beta(W+\triangle F)}}\right\rangle _{B},\label{Freendiff}
\end{equation}
in which $\triangle F=\triangle F_{\tau}$ is the free-energy difference
between the initial and final equilibrium states of the pulling, which
can be calculated from the Bennett acceptance ratio (BAR) method \cite{Bennett1976}.
Equation (5) can also be rearranged to give BAR formula
at the particular case where $t=\tau$:

\begin{equation}
e^{-\beta\triangle F}=\left\langle \frac{n_{F}e^{-\beta W}}{n_{F}+n_{B}e^{-\beta(W-\triangle F)}}\right\rangle _{F}+\left\langle \frac{n_{B}e^{\beta W}}{n_{F}+n_{B}e^{\beta(W+\triangle F)}}\right\rangle _{B}.\label{BAR}
\end{equation}

\paragraph{PMF as a Function of an Arbitrary CV - Reweighting Scheme
for Bi-directional Pullings}
The PMF computed as a function of the steered CV does not necessarily
provide a good picture of the investigated transition, as the steered
CV is not guaranteed to properly distinguish all the relevant states.
Moreover, in many cases it is instructive to look at the same result
from a different perspective, i.e. computing the PMF as a function
of a different, \emph{a posteriori} chosen CV. Such task can be performed
by employing a reweighting scheme. Suitable schemes have been proposed
for other kinds of non-equilibrium simulations including metadynamics
\cite{Bonomi2009}. For SMD simulations, a reweighting algorithm for
uni-directional pullings was introduced by some of us in a recent
work \cite{ColizziBussi2012}. Here we extend this scheme to bi-directional
pullings. As introduced in ref \cite{ColizziBussi2012}, at each frame
along the $i^{\textrm{th}}$ trajectory, a so-called \textit{weighting
factor} can be derived as:

\begin{equation}
w_{i}(t)\propto{\displaystyle \frac{e^{-\beta(W_{i}(t)-\triangle F_{t})}}{\sum_{t'}e^{-\beta(V(z_{t},t')-\triangle F_{t'})}}},\label{weight}
\end{equation}
where $t$ denotes the time frame; $W_{i}(t)$ is the work at time
$t$ done on the $i^{\textrm{th}}$ pulling; $V(z_{t},t')=k/2[z_{t}-(z_{0}+vt')]^{2}$
is the harmonic potential with a spring constant $k$ and the center
of oscillation at $z_{0}+vt'$ acting on the CV at time $t$ whose
value is $z_{t}$; $\triangle F_{t}$ is a normalization factor that
adopts the value of the free-energy difference between the equilibrium
state at time $t$ and the initial equilibrium state and can be calculated
as in Equation (5).

For bi-directional pulling, the weighting factor for both forward
and backward trajectories can be written as:

\begin{equation}
w_{i}^{F}(t)\propto{\displaystyle \frac{e^{-\beta(W_{i}(t)-\triangle F_{t})}}{\sum_{t'}e^{-\beta(V(z_{t},t')-\triangle F_{t'})}}},\label{weight-forward}
\end{equation}

\begin{equation}
w_{i}^{B}(t)\propto{\displaystyle \frac{e^{-\beta(W_{i}(t)-\triangle F_{\tau-t})}}{\sum_{t'}e^{-\beta(V(z_{t},t')-\triangle F_{t'})}}}.\label{weight-backward}
\end{equation}
Note that $\triangle F_{t}$ has been adjusted to $\triangle F_{\tau-t}$
in Equation (9) since the backward process
is actually a reverse of the forward one.

We notice that Equations (8) and (9)
do not fix the relative weight of forward and backward trajectories.
In the same spirit as in ref. \cite{MinhAdib2008} this can be optimally
done using the weighted-histogram analysis method (WHAM) \cite{Kumar1992,SouailleRoux2001},
which provides the optimal relative weight. Therefore, the weighting
factors in Equations (8) and (9)
can be rewritten as: 

\begin{equation}
w_{i}^{F}(t)\propto{\displaystyle \frac{e^{-\beta(W_{i}(t)-\triangle F_{t})}}{\sum_{t'}e^{-\beta(V(z_{t},t')-\triangle F_{t'})}}\times\frac{1}{n_{F}+n_{B}e^{-\beta(W_{i}(\tau)-\triangle F)}}},\label{weight-forward-mod}
\end{equation}

\begin{equation}
w_{i}^{B}(t)\propto{\displaystyle \frac{e^{-\beta(W_{i}(t)-\triangle F_{\tau-t})}}{\sum_{t'}e^{-\beta(V(z_{t},t')-\triangle F_{t'})}}}\times\frac{e^{-\beta\triangle F}}{n_{B}+n_{F}e^{-\beta(W_{i}(\tau)+\triangle F)}}.\label{weight-backward-mod}
\end{equation}

Equations (10) and (11)
are the combinations of the equations derived in \cite{MinhAdib2008,ColizziBussi2012}
and provide a weight to each of the sampled configurations. Based
on these weights, the PMF can be estimated as a function of
any \textit{a posteriori} chosen CV $\bar{z}$ as:

\begin{equation}
G(\bar{z})=-k_{B}T\log\sum_{t}\sum_{i}(w_{i}^{F}(t)+w_{i}^{B}(t))\delta(\bar{z}-\bar{z}_{t}).\label{freen-reweight}
\end{equation}

Our reweighting scheme (Equation (12)) can be rearranged
to give an identical expression to the Minh-Adib bi-directional PMF
estimator (Equation (4)) when $\bar{z}\equiv z$. However,
it is more general as it allows estimating the PMF as a function of
a different, \emph{a posteriori} chosen CV. Applying this algorithm,
one can be flexible in choosing the appropriate CVs for different
purposes.

\subsection{Computational Details}
All standard MD simulations were performed using GROMACS 4.5.5 \cite{GROMACS2008}.
A truncated octahedral box of explicit water at large size was used
to ensure that TAR and L22 can reach the center-to-center distance
of at least 40 {\AA{}}. The box turned out to contain 11,780 water
molecules. An excess ion concentration of 10 mM was set in all simulations
to reproduce the experimental conditions \cite{Davidson2009}, which
resulted in 23 K$^{+}$ cations and 2 Cl$^{-}$ anions. We employed
TIP3P model \cite{Jorgensen1983} for water, AMBER ff99SB-ILDN force
field \cite{Lindorff2010} for L22, and ff99SB-ILDN
with parmbsc0 reparametrization \cite{Perez2007} for TAR. When using
the standard ff99SB-ILDN force field for K$^{+}$ and Cl$^{-}$ ions
at the ion concentration of 150 mM, we experienced the growing of
ion crystallization after the first 5 ns of MD simulation (data not
shown). In fact, the AMBER ff9X force fields, i.e. ff94 \cite{Cornell1995},
ff98 \cite{CheathamIII1999}, and ff99 \cite{Wang2000,Hornak2006}
and their variants, have been reported to facilitate the ion crystallization
due to the incorrect parametrizations which cause the imbalance between
cation-anion interactions \cite{SavelyevPapoian2007,Auffinger2007,ChenPappu}.
Therefore, the ff99SB-ILDN force field corrected with new ions' reparametrization
by Joung and Cheatham \cite{JoungCheatham2008} was used for the K$^{+}$
and Cl$^{-}$ ions in our simulations. In all simulations, temperature
was kept constant at 300 K using the velocity rescaling algorithm
\cite{Bussi2007}. When indicated, pressure was kept constant at 1
atm using a Parrinello-Rahman barostat \cite{ParrinelloRahman1981}.

A total of 4.2 microseconds of simulations were performed in a bi-directional
steering scheme. Hereafter we refer to the unbinding direction as
\textit{forward} SMD and the binding direction as \textit{backward}
SMD.

Our forward SMD scheme included:
\begin{description}
\item [{\textmd{\textit{(1f)}}}] 20 ns of constant-pressure MD simulation
starting from the NMR structure of L22-TAR complex. An average DHEN
($-140$ kJ/mol), and its standard deviation ($\sigma=3$ kJ/mol)
were calculated from this equilibrium simulation.
\item [{\textmd{\textit{(2f)}}}] 64 ns of CV-restrained simulation in which
the value of the DHEN was restrained to the value of $-140$ kJ/mol.
A spring constant $k=0.2$ kJ mol$^{-1}$ nm$^{-2}$$\approx k_{B}T/\sigma^{2}$
was used here and in the following SMD simulations.
\item [{\textmd{\textit{(3f)}}}] Configurations were then extracted every
1 ns and used as initial structures for 64 forward (unbinding) SMD
simulations (25 ns each), in which the DHEN was pulled at a constant
velocity from the value of $-140$ kJ/mol to $-30$ kJ/mol. This target
value has been chosen large enough so that the two molecules are completely
separated. Indeed, among 64 final structures of unbinding SMD simulations,
the smallest center-to-center distance between L22 and TAR is about
32 Å while the smallest distance between an L22 atom and a TAR atom
is about 7 Å.
\end{description}
Our backward SMD scheme consisted of the following steps:
\begin{description}
\item [{\textmd{\textit{(1b)}}}] Each of the structures obtained at the
end of the forward SMD was equilibrated for 1 ns, restraining the
DHEN at $-30$ kJ/mol.
\item [{\textmd{\textit{(2b)}}}] A random configuration was then extracted
from each CV-restrained simulation and used as the starting structure
for another set of 64 backward (binding) SMD (25 ns each), in which
the DHEN CV was pulled from $-30$ kJ/mol back to $-140$ kJ/mol with
the same velocity and spring constant as in the forward SMD simulations.
\end{description}
Among 64 bound configurations at the end of the backward SMD simulations,
we found two dominant classes in which L22 binds to the TAR's upper
major groove, i.e. the same binding pocket observed in NMR studies.
To further quantify the difference between these two classes, for
each class, we selected the structure associated with the lowest steering
work. We then repeated the same forward SMD procedure as described
above, which, for each selected structure, included \textit{(i)} 30
ns of MD equilibration for step \textit{(1f)}, \textit{(ii)} 12 ns
of CV-restraint MD for step \textit{(2f)} at the same DHEN value ($-140$
kJ/mol ) and spring constant (0.2 kJ mol$^{-1}$ nm$^{-2}$), and
\textit{(iii)} 12 forward SMD simulations (25 ns each) for step \textit{(3f)}.

SMD and CV-restrained simulations were performed using our modified
version of PLUMED 1.3 integrated with GROMACS 4.5.5. In all these
simulations, Equation (3) was used to determine the DHEN
CV. All atoms of L22 and TAR are involved in the sets $A$ and $B$
respectively. Atomic charges were extracted from the ff99SB-ILDN force
field. The value of 80 was chosen for the dielectric constant of water
$\epsilon_{w}$, which, together with the ionic strength $I=10$ mM,
resulted in the Debye-H\"uckel parameter $\kappa\simeq0.033$ {\AA{}}$^{-1}$.


\section{Results}

\subsection{PMF Reconstruction from Bi-directional SMD Simulations}
\begin{figure}
\includegraphics[width=0.6\textwidth]{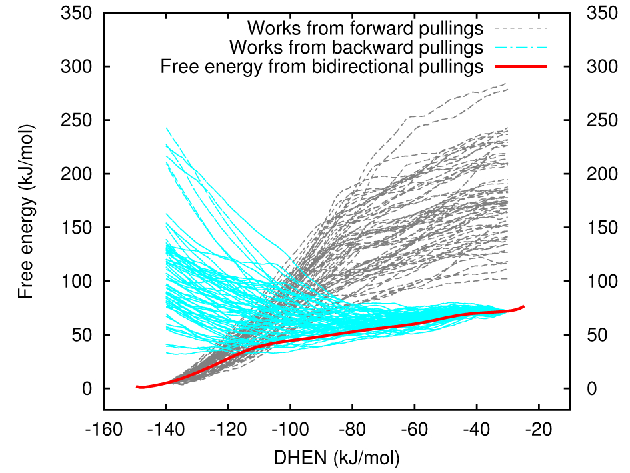}
\caption{Reconstruction of PMF as a function
of DHEN CV (solid red line). The works from forward and backward pullings
are also shown (dotted gray and cyan lines respectively). Backward
works are shifted by $\Delta F$ as estimated from Equation (6).}
\label{FreeEnergyMinhAdib}
\end{figure}

\begin{figure}
\includegraphics[width=0.6\textwidth]{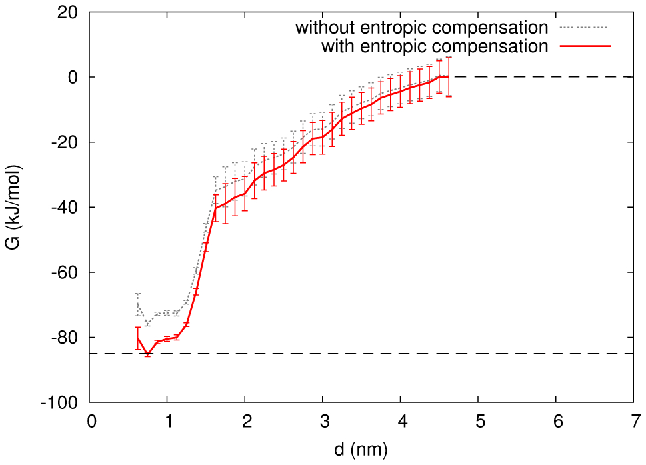}
\caption{PMF as a function of the geometric center-to-center distance
($d$) obtained by our proposed reweighting scheme (Equation (12))
without any entropic compensation (dotted gray line) and with an entropic
compensation of $2k_{B}T\log d$ (red solid line). The plots have
been shifted so that PMF profiles are aligned at $d=4.5$ nm. The
projection of PMF onto distance allows defining both
bound and unbound states. The free-energy difference between these
states is approximately $85\pm5$ kJ/mol. Errors are calculated using
the bootstrapping algorithm.}
\label{FreeEnergyReweightDistance}
\end{figure}

The PMF as a function of DHEN was reconstructed from bi-directional
pullings using the Minh-Adib PMF estimator as in Equation (4)
(see Figure \ref{FreeEnergyMinhAdib}). However, within this framework,
it is difficult to quantify the entropic contribution to the free
energy arising from the larger conformational space available as the
intermolecular distance increases, which is expected to diverge for
the unbound state. We then apply the reweighting scheme (Equation (12)) to compute the PMF as
a function of the distance between the centers of mass of the two
molecules (see Figure \ref{FreeEnergyReweightDistance}). This entropic contribution
can be easily evaluated and added to the PMF, and hence allows estimating
the free-energy difference between the bound and unbound states, which
turns out to be approximately $85\pm5$ kJ/mol. This value is larger
than that obtained from experiment, which was approximately 52 kJ/mol
\cite{Davidson2009}. In the Section Discussions, we prove that this discrepancy
does not come from sampling problems, and it is presumably due to the inaccuracy
of the force fields.

\subsection{Binding Poses from Backward SMD Simulations}
\begin{figure}
\includegraphics[width=0.6\textwidth]{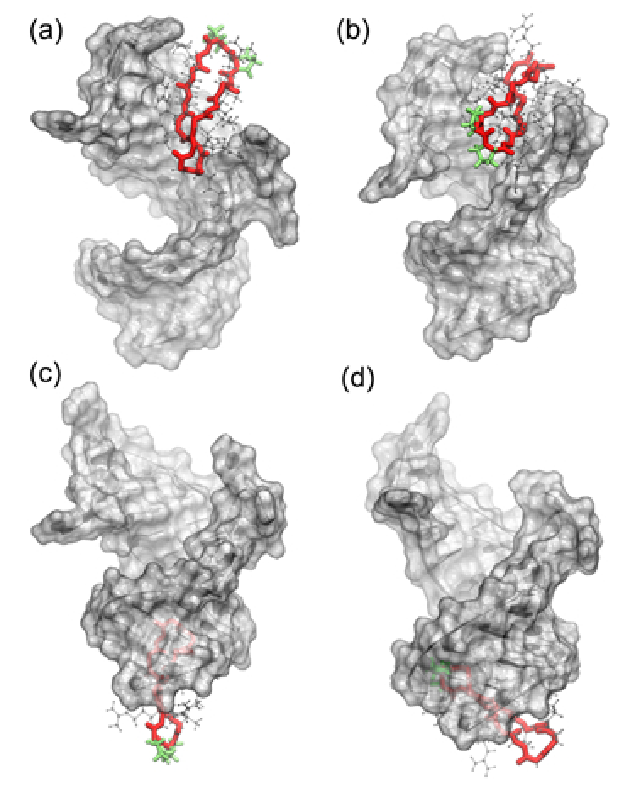}
\caption{Representatives of four dominant binding poses obtained from 64 binding
SMD simulations including (a) upper-major-groove binding pose with
the $^{\textrm{L}}$Pro$-$$^{\textrm{D}}$Pro template, colored in
green, points outward TAR (i.e. pose \textbf{(1)} as classified in
Table \ref{TabClassPose}), (b) upper-major-groove binding pose with
$^{\textrm{L}}$Pro$-$$^{\textrm{D}}$Pro points outward TAR (pose
\textbf{(2)}), (c) lower-major-groove binding pose with $^{\textrm{L}}$Pro$-$$^{\textrm{D}}$Pro
points outward TAR (pose \textbf{(4)}), and (d) lower-major-groove
binding pose with $^{\textrm{L}}$Pro$-$$^{\textrm{D}}$Pro points
inward TAR (pose \textbf{(5)}). Pose \textbf{(2)} is close the NMR
structure (see Figure \ref{Structures}(c)). }
\label{MajorPoses}
\end{figure}

A total of 64 binding (backward) SMD simulations, in which the DHEN
CV was pulled from $-30$ kJ/mol to $-140$ kJ/mol, all ended up with
L22-TAR complexes. The binding poses of L22 to TAR can be classified
as followed (see Table \ref{TabClassPose})
\begin{description}
\item [{\textmd{\textit{(i)}}}] L22 binds to TAR at the major groove in
51 complexes (80\%), among which 33 complexes (52\%) can be classified
as upper-major-groove binding (i.e. the same binding pocket as of
the Tat protein and as observed in NMR experiment \cite{Davidson2009}),
12 complexes (19\%) feature lower-major-groove binding, and 6 complexes
(9\%) have L22 bind to TAR at the region lying between the upper and
lower major groove (referred to as middle major groove hereafter).
\item [{\textmd{\textit{(ii)}}}] L22 binds to TAR at the minor groove in
10 complexes (16\%), among which there are 4 complexes (7\%) classified
as upper-minor-groove binding and 6 complexes (9\%) classified as
lower-minor-groove binding.
\item [{\textmd{\textit{(iii)}}}] 3 complexes (4\%) do not belong to any
of the above categories.
\end{description}
In the upper-major-groove binding poses, we found 20 complexes (31\%)
in which the $^{\textrm{L}}$Pro$-$$^{\textrm{D}}$Pro template of
L22 points outward TAR (i.e. pose \textbf{(1)} in Table \ref{TabClassPose}
and Figure \ref{MajorPoses}(a) and 8 complexes (13\%) in which the
$^{\textrm{L}}$Pro$-$$^{\textrm{D}}$Pro template points inward
TAR (pose \textbf{(2)} in Table \ref{TabClassPose} and Figure \ref{MajorPoses}(b).
Similarly, in the lower-major-groove binding, we also found 6 complexes
(9\%) with the $^{\textrm{L}}$Pro$-$$^{\textrm{D}}$Pro template
pointing outward (pose \textbf{(4)} in Table \ref{TabClassPose} and
Figure \ref{MajorPoses}(c) and 4 complexes (7\%) with the $^{\textrm{L}}$Pro$-$$^{\textrm{D}}$Pro
template pointing inward (pose \textbf{(5)} in Table \ref{TabClassPose}
and Figure \ref{MajorPoses}(d). Pose \textbf{(2)} represents the same
binding pose as observed in the NMR experiment \cite{Davidson2009}.
However, it only appears as the second-most probable pose in our simulations.
The most probable pose has the same binding pocket but with $^{\textrm{L}}$Pro$-$$^{\textrm{D}}$Pro
pointing to the opposite direction.

A more appropriate assessment of relative pose stability can be done
by considering the works performed in the pulling simulations. Among
the five simulations with lowest pulling works, only the one with
the second-lowest work ended up in pose \textbf{(3)} (see Table \ref{TabClassPose}).
The rest of them are classified as pose \textbf{(1)}. The sixth-lowest-work
process results in a complex of pose \textbf{(2)}.

\begin{table}
\begin{tabular}{c c c c c c c c c c}
\hline
\hline
\multicolumn{7}{c}{Major-groove binding (80\%)} & \multicolumn{2}{c}{Minor-groove binding (16\%)} & \multirow{3}{*}{Others}\tabularnewline
\cline{1-9}
\multicolumn{3}{c}{Upper (52\%)} & \multicolumn{3}{c}{Lower (19\%)} & \multirow{2}{*}{Middle} & \multirow{2}{*}{Upper} & \multirow{2}{*}{Lower} & \tabularnewline
\cline{1-6}
\textbf{(1)} & \textbf{(2)} & \textbf{(3)} & \textbf{(4)} & \textbf{(5)} & \textbf{(6)} &  &  &  & \tabularnewline
\hline
31\% & 13\% & 8\% & 9\% & 7\% & 3\% & 9\% & 7\% & 9\% & 4\%\tabularnewline
\hline
\hline
\end{tabular}
\caption{Occurrence of L22-TAR binding poses obtained from 64 binding SMD simulations.
To better classify the binding poses in the major groove, we subdivide
the upper-groove and lower-groove binding poses into smaller groups.
For the upper groove, \textbf{(1)} represents the pose in which the
$^{\textrm{L}}$Pro$-$$^{\textrm{D}}$Pro template of L22 points
outward TAR; \textbf{(2)} denotes the pose with the $^{\textrm{L}}$Pro$-$$^{\textrm{D}}$Pro
template pointing inward TAR; and \textbf{(3)} contains the rest of
the upper-major-groove-binding complexes. Similarly, poses \textbf{(4)},
\textbf{(5)}, and \textbf{(6)} respectively represents the same classification
criteria for the lower-major-groove binding. It is noteworthy that
pose \textbf{(2)} is the binding pose observed in the NMR experiment.}
\label{TabClassPose}
\end{table}

\subsection{Quantitative Comparison in the Stability of the Two Dominant Upper-Major-Groove
Binding Poses\label{sub:Quantitative-comparison-in}}
\begin{figure}
\includegraphics[width=0.6\textwidth]{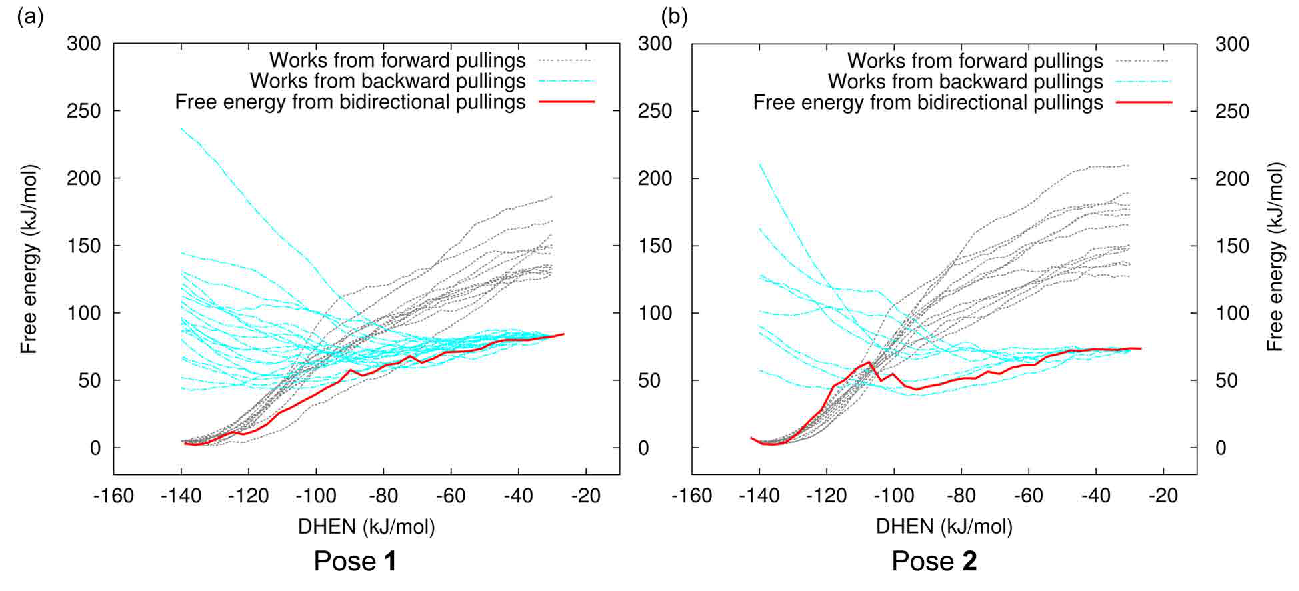}
\caption{Reconstruction of PMF in pose \textbf{(1)}
(a) pose \textbf{(2)} (b) as a function of DHEN CV (solid red line).
The works from forward and backward pullings are also shown (gray
lines and cyan lines respectively).}
\label{Pose1Pose2MinhAdib}
\end{figure}

As discussed in the previous Section, pose \textbf{(1)} in Figure
\ref{MajorPoses} is not only the most probable pose but also the
pose for which the lowest work is performed. In pose \textbf{(2)},
the ligand occupy the same binding pocket in a different orientation,
which is consistent with that obtained from NMR data. For a quantitative
comparison between pose \textbf{(1)} and pose \textbf{(2)}, we performed
12 unbinding SMD simulations starting from the complex obtained by
the lowest work in each pose (i.e. the globally lowest work in case
of pose \textbf{(1)} and the sixth-lowest work in case of pose \textbf{(2)}).
We then combined these 12 unbinding simulations of each pose with
a selected set of the previous binding simulations which ended on
the same pose (i.e. 20 binding simulations resulting in pose \textbf{(1)}
and 8 simulations resulting in pose \textbf{(2)}). Equation (4)
was then used to calculate the PMF as a function of the restrained
DHEN CV based on such combination of each pose (see Figure \ref{Pose1Pose2MinhAdib}).
Especially, this equation is applicable even to different numbers
of forward and backward trajectories. The free-energy difference between
the two end states associated with pose \textbf{(1)} is larger than
that of pose \textbf{(2)} (79 versus 69 kJ/mol respectively), thus
pose \textbf{(1)} can be considered more stable than pose \textbf{(2)}.

To verify the robustness of the comparison, we performed a bootstrapping
algorithm by repeating 500 times of solving the BAR equation (6)
(which is a special case of Equation (5) when one considers
only the end states) on the randomly chosen half-set, i.e. 6 unbinding
and 10 binding simulations for pose \textbf{(1)} and 6 unbinding and
4 binding simulations for pose \textbf{(2)}. The resulted free energy
difference from 500 BAR calculations for each pose can be found in
Supporting Information. The average values of free-energy difference
are $75\pm6$ and $66\pm6$ kJ/mol for pose \textbf{(1)} and pose
\textbf{(2)} respectively. Despite random choice of the half-set of
simulations to be involved in BAR calculations, pose \textbf{(1)}
consistently shows a higher stability than pose \textbf{(2)}.

\begin{figure}
\includegraphics[width=0.6\textwidth]{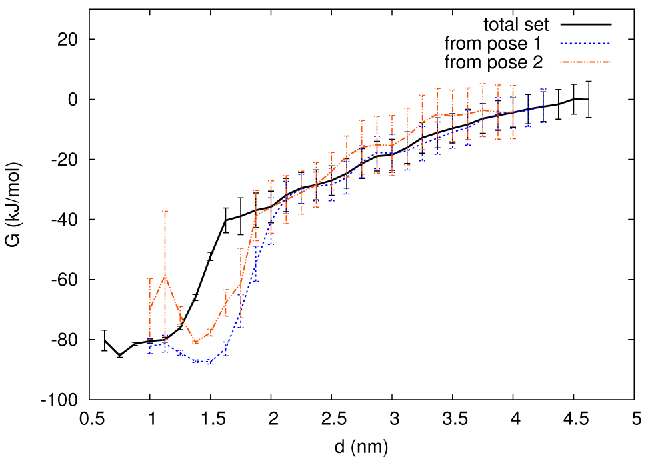}
\caption{PMF as a function of the center-to-center distance obtained
by the reweighting scheme performed on the whole set of simulations
starting from the NMR structure (black solid line) and the set of
simulations coming from and to pose \textbf{(1)} (blue dotted line)
and pose \textbf{(2)} (red dashed line). Free-energy difference between
the bound and unbound states of pose \textbf{(1)} is larger than those
of the NMR structure and pose \textbf{(2)}.}
\label{FreeEnergyReweightDistanceThree}
\end{figure}

Applying the proposed reweighting scheme on the center-to-center distance
CV, we again found that the free-energy differences as functions of
distance of both poses \textbf{(1)} is larger than that of pose \textbf{(2)
}and the NMR structure (see Figure \ref{FreeEnergyReweightDistanceThree}).

\section{Discussions\label{sec:Discussions}}

\subsection*{Effectivity, Computational Efficiency, and Generality of the Proposed
Electrostatic-Based Collective Variable}
Our proposed DHEN CV (Equation (3)) includes only the intermolecular
electrostatic interactions thus does not contain the noise coming
from intramolecular interactions or interactions with the ionic solvent.
Our CV is a function of not only atomic coordinates but also ionic
strength, temperature, solvent dielectric constant, and atomic charges
accessible from the force fields. Therefore, it is expected to be
more ``selective'' and effective than the conventional center-to-center
distance CV in describing RNA/peptide binding/unbinding processes.
Indeed, the distance CV may disfavor the right complex formation by
not taking into account the charge-charge interactions, an important
driving force in most RNA/peptide recognition events.

Besides the parameters that are trivially determined from the simulation
setups, our proposed CV does not requires any other extra parameters
that need to be updated during the simulations. This is the computational
advantage of this CV.

In addition, our proposed CV has a general formalism that, in theory,
is applicable for any binding event that is driven by electrostatics,
which is the case of most nucleic acid/ligand bindings.

\subsection*{Bi-directional Steering Outperforms Uni-directional Steering}
Steering involves out-of-equilibrium processes. Although the uni-directional
Jarzynski's equality \cite{Jarzynski1997} and the Hummer-Szabo PMF
estimator \cite{HummerSzabo2001} allow reconstructing the free energy
from nonequilibrium works, the resulted free energy is strongly dominated
by the low work values due to the behavior of the exponential average.
These low work values are associated with the trajectories of the
``rare'' events which are poorly sampled. This poses a convergence
challenge to uni-directional SMD simulations. Indeed, the more a system
departs from its initial equilibrium state, the more uni-directional
estimators tend to overestimate the free energy change.

The bi-directional estimators such as the one introduced by Minh-Adib
\cite{MinhAdib2008} and our proposed reweighting scheme provide an
optimal combination of the works from both forward and backward processes.
When the forward and backward processes are properly combined, the
overestimation of free energy in both directions are then ``averaged''
out. Bi-directional estimators hence outperform uni-directional ones.

\subsection*{Verification of the Statistical Accuracy}
The bootstrapping procedure of analyzing a selected subset of the
binding and unbinding trajectories allows us to assess the statistical
accuracy of the results. Indeed, bi-directional pulling protocols
can still exhibit large statistical errors when backward pullings
do not properly bring the system to the correct starting point. However,
by performing the pulling out of the actually reached bound pose, we can guarantee that their
stability is fairly evaluated.

\subsection*{\textquotedblleft{}Blind\textquotedblright{} Prediction of the Binding
Pocket}
By employing bi-directional SMD simulations with our proposed DHEN
CV, we were able to find the binding pocket in agreement with the
experimental structure, i.e. the upper major groove. Although the
L22-TAR binding pose and its structural properties were well characterized
in experiments, we did not use any of these experimental observations
to bias or constraint the simulations. Indeed, our pullings were performed
in a ``blind'' manner, in which L22 is not constrained to bind to
the known NMR binding pocket but rather free to decide its encounter
paths guided by the electrostatic interactions. In such manner, we
are still able to find two dominant binding poses, in both of which
L22 binds to the correct pocket as observed by experiments.
Figure S3 in Supporting Information shows the
positive correlation between the total works and the last values
of DHEN in 64 backward simulations. It is clear that the DHEN CV points
to the upper major groove as the best binding site. However, we also observe
that the total work is a better estimator of the reliability of the final
conformation reached by a steered simulation since {\it (i)} it takes into account
the entire binding process and {\it (ii)} it has a rigorous foundation which,
in the limit of an infinite number of pulling simulations, is independent
of the way the DHEN is estimated. We recalculated the DHEN using 100 mM as the
ionic strength  for the final conformations resulted from the backward steered
trajectories. A positive correlation between these new DHEN values and the total
works (see Figure S4 in Supporting Information) suggests that this method would
be able to distinguish the two grooves also at higher ion concentrations.
These results not only confirms the assumption that electrostatics plays an important
role in L22-TAR binding but also strongly justifies the use of DHEN,
an approximation of the electrostatic interaction free energy, as
a CV in accelerated simulations.

TAR is an RNA containing 29 nucleotides featuring two stems, a bulge,
and an apical loop. The apical loop partially closes the access to
the upper major groove associated with the upper stem, which is also
the binding pocket of Tat. Any designed molecule able to bind to TAR
at this pocket is a promising candidate for HIV-transcription inhibition.
L22 is a ligand with a rigid $\beta-$hairpin backbone and long side-chains
(i.e. mostly composed of Arginine side-chains), which make it difficult
to navigate and end up inside a partially closed pocket. Interestingly,
in experiments, L22 was reported to bind and fit completely in this
pocket. We were able to reproduce such a non-trivial binding mode
only by using SMD simulations pulling on an electrostatic-potential-energy-based
CV without any further guidance from experiments. The self-guiding
feature of this methodology is very important since it is applicable
even when experimental information in unavailable, which is the case
of most computational drug design efforts. This feature could further
help designing a robust and automatic computational strategy for studying
biomolecular complexes and hence help cutting experimental expenses.

\subsection*{Predicted Affinity}
As we discussed above, we are confident in the statistical accuracy
of our calculations. Moreover, contributions such as the conformational entropy loss upon binding
are implicitly taken into account in our approach. Thus, the discrepancy between
our estimate of the affinity and the one reported in ref. \cite{Davidson2009}
should be ascribed to other causes such as the difference between
the anion type we used (Cl$^{-}$) and the one used in the experiments
(mixture of HPO$_{4}^{2-}$ and H$_{2}$PO$_{4}^{-}$); the not so
large simulation box and the resulting periodic boundary artifacts
which are even more amplified in smaller simulation boxes; and finally, force field inaccuracies.
These latter inaccuracies could be likely
associated with \textit{(i)} the challenging description of the multi-degree-of-freedom
sugar-phosphate backbone using a constant-point-charge model \cite{Reynolds1992,Cieplak1995},
\textit{(ii)} the difficulties in describing the RNA non-canonical
structural elements \cite{Orozco2008,Banas2010,Yildirim2011,Denning2011},
i.e. the bulge and hairpin loop in our case, \textit{(iii)} subtle
force-field dependence on ionic strength and types \cite{Besseova2009,Noy2009},
and most importantly \textit{(iv)} the well-known inaccuracies of
the non-polarizable force fields in describing the electrostatic interaction
between the RNA and the anions (i.e, Cl$^{-}$ions) as well as the
strong electrostatic interaction between the highly polarizable phosphodiester
moiety of RNA and the positively charged atoms \cite{McDowell2007,FulleGohlke2010}
including the cations K$^{+}$ and those of the peptidic ligand L22
in our case.

\subsection*{Sufficiency of DHEN as a CV}
A physics-based CV is expected to outperform a geometrical CV in guiding the
transition process. However, it is not guaranteed that our DHEN CV is sufficient,
since it may not distinguish some degenerated conformations.
In this work, the effect of this drawback is reduced by performing
multiple steered simulations to sample other possibly hidden relevant degrees of freedom
with a brute-force approach.
Subsequently, the obtained conformations can be interpreted with their correct statistical weights
using our reweighting scheme.
In other simulation frameworks, such as metadynamics or umbrella sampling with weighted histogram analysis method,
orthogonal CVs could be biased simultaneously with DHEN CV to alleviate this issue.

\section{Conclusions}
Motivated by the idea of using potential energy of a system as a collective
variable and based on the observation that binding between an RNA
and a positively charged ligand is driven by electrostatic interaction,
we have proposed an electrostatic-based collective variable that is
an approximation of the intermolecular electrostatic
component of the free energy. Our proposed collective variable is
a physics-based indicator which is computationally efficient and simple
to be integrated with atomistic simulations%
\bibnote{Our modifications to the PLUMED software are available on request%
}. By performing bi-directional steered molecular dynamics on this
collective variable, and taking advantage of a novel reweighting scheme,
we provide quantitative insight on the binding energetics of L22-TAR
complex, and blindly reproduce the correct binding pocket and pose
as observed in NMR experiments. This shows an efficient self-guiding
feature of our simulation protocol, which is extremely important for
drug design when experimental structures are not available. A statistical
validation of our results allows the overestimation of the binding
affinity to be ascribed to the well-known force-field deficiencies
and/or to other differences between the computational and experimental
setups. We foresee the application of this technique in studying other electrostatics-driven
processes, such as assembly of complexes of charged molecules including ligand/nucleic acid,
protein/nucleic acid, protein/protein interactions. The DHEN collective variable can also be used within the framework of other biasing techniques
such as umbrella sampling, adaptive biasing force, and metadynamics,
and with temperature-based schemes. As a possible extension, more
advanced estimates of the electrostatic interaction free-energy, such as those based on
Generalized Born model \cite{Bashford2000}, could be used as collective variables for biased sampling.

\acknowledgement
We acknowledge the CINECA Award no. HP10AQXC5N for the availability
of high performance computing resources. G.B. acknowledges the European
Research Council for funding through the Starting Grant S-RNA-S (no.
306662).
G.V. was supported by NSF grant MCB-1050702.
This work has been supported by the BANDO AIDS Grant from
the Italian Government for ``Targeting HIV transcription
to control infection and to purge post-integrative latency.''

\section*{Supporting Information Available}
A detailed derivation of the Debye-H\"uckel-free-energy collective variable is available in Supporting Information.
This material is available free of charge via the Internet at {\tt http://pubs.acs.org/}.

\bibliography{bibTARSMD}

\providecommand*\mcitethebibliography{\thebibliography}
\csname @ifundefined\endcsname{endmcitethebibliography}
  {\let\endmcitethebibliography\endthebibliography}{}
\begin{mcitethebibliography}{88}
\providecommand*\natexlab[1]{#1}
\providecommand*\mciteSetBstSublistMode[1]{}
\providecommand*\mciteSetBstMaxWidthForm[2]{}
\providecommand*\mciteBstWouldAddEndPuncttrue
  {\def\EndOfBibitem{\unskip.}}
\providecommand*\mciteBstWouldAddEndPunctfalse
  {\let\EndOfBibitem\relax}
\providecommand*\mciteSetBstMidEndSepPunct[3]{}
\providecommand*\mciteSetBstSublistLabelBeginEnd[3]{}
\providecommand*\EndOfBibitem{}
\mciteSetBstSublistMode{f}
\mciteSetBstMaxWidthForm{subitem}{(\alph{mcitesubitemcount})}
\mciteSetBstSublistLabelBeginEnd
  {\mcitemaxwidthsubitemform\space}
  {\relax}
  {\relax}

\bibitem[Froeyen and Herdewijn(2002)Froeyen, and
  Herdewijn]{FroeyenHerdewijn2002}
Froeyen,~M.; Herdewijn,~P. {RNA} as a Target for Drug Design, the Example of
  {T}at-{TAR} Interaction. \emph{Curr. Top. Med. Chem.} \textbf{2002},
  \emph{2}, 1123--1145\relax
\mciteBstWouldAddEndPuncttrue
\mciteSetBstMidEndSepPunct{\mcitedefaultmidpunct}
{\mcitedefaultendpunct}{\mcitedefaultseppunct}\relax
\EndOfBibitem
\bibitem[Zacharias(2003)]{Zacharias2003}
Zacharias,~M. Perspectives of Drug Design that Targets {RNA}. \emph{Curr. Med.
  Chem.: Anti-Infect. Agents.} \textbf{2003}, \emph{2}, 161--172\relax
\mciteBstWouldAddEndPuncttrue
\mciteSetBstMidEndSepPunct{\mcitedefaultmidpunct}
{\mcitedefaultendpunct}{\mcitedefaultseppunct}\relax
\EndOfBibitem
\bibitem[Vicens(2009)]{Vicens2009}
Vicens,~Q. {RNA}'s coming of age as a rug target. \emph{J. Incl. Phenom.
  Macrocycl. Chem.} \textbf{2009}, \emph{65}, 171--188\relax
\mciteBstWouldAddEndPuncttrue
\mciteSetBstMidEndSepPunct{\mcitedefaultmidpunct}
{\mcitedefaultendpunct}{\mcitedefaultseppunct}\relax
\EndOfBibitem
\bibitem[Puglisi et~al.(1993)Puglisi, Chen, Frankelti, and
  Williamson]{Puglisi1993}
Puglisi,~J.; Chen,~L.; Frankelti,~A.; Williamson,~J. Role of {RNA} Structure in
  {A}rginine Recognition of {TAR} {RNA}. \emph{Proc. Natl. Acad. Sci. U.S.A.}
  \textbf{1993}, \emph{90}, 3680--3684\relax
\mciteBstWouldAddEndPuncttrue
\mciteSetBstMidEndSepPunct{\mcitedefaultmidpunct}
{\mcitedefaultendpunct}{\mcitedefaultseppunct}\relax
\EndOfBibitem
\bibitem[Auffinger and Westhof(2000)Auffinger, and
  Westhof]{AuffingerWesthof2000}
Auffinger,~P.; Westhof,~E. Water and Ion Binding around {RNA} and {DNA}
  ({C},{G}) Oligomers. \emph{J. Mol. Biol.} \textbf{2000}, \emph{300},
  1133--1131\relax
\mciteBstWouldAddEndPuncttrue
\mciteSetBstMidEndSepPunct{\mcitedefaultmidpunct}
{\mcitedefaultendpunct}{\mcitedefaultseppunct}\relax
\EndOfBibitem
\bibitem[Hermann(2002)]{Hermann2002}
Hermann,~T. Rational ligand design for {RNA}: the role of static structure and
  conformational f{}lexibility in target recognition. \emph{Biochimie}
  \textbf{2002}, \emph{84}, 869--875\relax
\mciteBstWouldAddEndPuncttrue
\mciteSetBstMidEndSepPunct{\mcitedefaultmidpunct}
{\mcitedefaultendpunct}{\mcitedefaultseppunct}\relax
\EndOfBibitem
\bibitem[Hermann and Tor(2005)Hermann, and Tor]{HermannTor2005}
Hermann,~T.; Tor,~Y. {RNA} as a Target for Small-molecule Therapeutics.
  \emph{Expert Opin. Ther. Patents} \textbf{2005}, \emph{15}, 49--62\relax
\mciteBstWouldAddEndPuncttrue
\mciteSetBstMidEndSepPunct{\mcitedefaultmidpunct}
{\mcitedefaultendpunct}{\mcitedefaultseppunct}\relax
\EndOfBibitem
\bibitem[Auffinger and Hashem(2007)Auffinger, and Hashem]{AuffingerHashem2007}
Auffinger,~P.; Hashem,~Y. Nucleic Acid Solvation: from Outside to Insight.
  \emph{Curr. Opin. Struc. Biol.} \textbf{2007}, \emph{17}, 325--333\relax
\mciteBstWouldAddEndPuncttrue
\mciteSetBstMidEndSepPunct{\mcitedefaultmidpunct}
{\mcitedefaultendpunct}{\mcitedefaultseppunct}\relax
\EndOfBibitem
\bibitem[Gallego and Varani(2001)Gallego, and Varani]{GallegoVarani2001}
Gallego,~J.; Varani,~G. Targeting {RNA} with small-molecule drugs: therapeutic
  promise and chemical challenges. \emph{Acc. Chem. Res.} \textbf{2001},
  \emph{34}, 836--843\relax
\mciteBstWouldAddEndPuncttrue
\mciteSetBstMidEndSepPunct{\mcitedefaultmidpunct}
{\mcitedefaultendpunct}{\mcitedefaultseppunct}\relax
\EndOfBibitem
\bibitem[Bosshard(2001)]{Bosshard2001}
Bosshard,~H. Molecular recognition by induced fit: how fit is the concept?
  \emph{News Physiol. Sci.} \textbf{2001}, \emph{16}, 171--173\relax
\mciteBstWouldAddEndPuncttrue
\mciteSetBstMidEndSepPunct{\mcitedefaultmidpunct}
{\mcitedefaultendpunct}{\mcitedefaultseppunct}\relax
\EndOfBibitem
\bibitem[Boehr et~al.(2009)Boehr, Nussinov, and Wright]{Boehr2009}
Boehr,~D.; Nussinov,~R.; Wright,~P. The role of dynamic conformational
  ensembles in biomolecular recognition. \emph{Nat. Chem. Biol.} \textbf{2009},
  \emph{5}, 789--796\relax
\mciteBstWouldAddEndPuncttrue
\mciteSetBstMidEndSepPunct{\mcitedefaultmidpunct}
{\mcitedefaultendpunct}{\mcitedefaultseppunct}\relax
\EndOfBibitem
\bibitem[Fulle and Gohlke(2010)Fulle, and Gohlke]{FulleGohlke2010}
Fulle,~S.; Gohlke,~H. Molecular Recognition of {RNA}: {C}hallenges for
  Modelling Interactions and Plasticity. \emph{J. Mol. Recognit.}
  \textbf{2010}, \emph{23}, 220--231\relax
\mciteBstWouldAddEndPuncttrue
\mciteSetBstMidEndSepPunct{\mcitedefaultmidpunct}
{\mcitedefaultendpunct}{\mcitedefaultseppunct}\relax
\EndOfBibitem
\bibitem[Leulliot and Varani(2001)Leulliot, and Varani]{LeulliotVarani2001}
Leulliot,~N.; Varani,~G. Current topics in {RNA}-protein recognition: control
  of specif{}icity and biological function through induced f{}it and
  conformational capture. \emph{Biochemistry} \textbf{2001}, \emph{40},
  7947--7956\relax
\mciteBstWouldAddEndPuncttrue
\mciteSetBstMidEndSepPunct{\mcitedefaultmidpunct}
{\mcitedefaultendpunct}{\mcitedefaultseppunct}\relax
\EndOfBibitem
\bibitem[Al-Hashimi(2005)]{AlHashimi2005}
Al-Hashimi,~H. Dynamics-based amplif{}ication of {RNA} function and its
  characterization by using {NMR} spectroscopy . \emph{Chem. Bio. Chem.}
  \textbf{2005}, \emph{6}, 1506--1519\relax
\mciteBstWouldAddEndPuncttrue
\mciteSetBstMidEndSepPunct{\mcitedefaultmidpunct}
{\mcitedefaultendpunct}{\mcitedefaultseppunct}\relax
\EndOfBibitem
\bibitem[Hamy et~al.(1997)Hamy, Felder, Heizmann, Lazdins, Aboul-Ela, Varani,
  Karn, and Klimkait]{Hamy1997}
Hamy,~F.; Felder,~E.; Heizmann,~G.; Lazdins,~J.; Aboul-Ela,~F.; Varani,~G.;
  Karn,~J.; Klimkait,~T. An inhibitor of the {T}at/{TAR} {RNA} interaction that
  effectively suppresses {HIV}-1 replication. \emph{Proc. Natl. Acad. Sci.
  U.S.A.} \textbf{1997}, \emph{94}, 3548--3553\relax
\mciteBstWouldAddEndPuncttrue
\mciteSetBstMidEndSepPunct{\mcitedefaultmidpunct}
{\mcitedefaultendpunct}{\mcitedefaultseppunct}\relax
\EndOfBibitem
\bibitem[Lipari and Szabo(1982)Lipari, and Szabo]{LipariSzabo1982-1}
Lipari,~G.; Szabo,~A. Model Free Approach to the Interpretation of {N}uclear
  {M}agnetic {R}esonance Relaxation in Macromolecules. 1. Theory and Range of
  Validity. \emph{J. Am. Chem. Soc.} \textbf{1982}, \emph{104},
  4546--4559\relax
\mciteBstWouldAddEndPuncttrue
\mciteSetBstMidEndSepPunct{\mcitedefaultmidpunct}
{\mcitedefaultendpunct}{\mcitedefaultseppunct}\relax
\EndOfBibitem
\bibitem[Lipari and Szabo(1982)Lipari, and Szabo]{LipariSzabo1982-2}
Lipari,~G.; Szabo,~A. Model Free Approach to the Interpretation of {N}uclear
  {M}agnetic {R}esonance Relaxation in Macromolecules. 2. Analysis of
  Experimental Results. \emph{J. Am. Chem. Soc.} \textbf{1982}, \emph{104},
  4559--4570\relax
\mciteBstWouldAddEndPuncttrue
\mciteSetBstMidEndSepPunct{\mcitedefaultmidpunct}
{\mcitedefaultendpunct}{\mcitedefaultseppunct}\relax
\EndOfBibitem
\bibitem[Setny et~al.(2012)Setny, Bahadur, and Zacharias]{Setny2012}
Setny,~P.; Bahadur,~R.; Zacharias,~M. Protein-{DNA} docking with a
  coarse-grained force {fi}eld. \emph{BMC Bioinformatics} \textbf{2012},
  \emph{13}, 228\relax
\mciteBstWouldAddEndPuncttrue
\mciteSetBstMidEndSepPunct{\mcitedefaultmidpunct}
{\mcitedefaultendpunct}{\mcitedefaultseppunct}\relax
\EndOfBibitem
\bibitem[Banavali and Roux(2002)Banavali, and Roux]{BanavaliRoux2002}
Banavali,~N.; Roux,~B. Atomic radii for continuum electrostatics calculations
  on nucleic acids. \emph{J. Phys. Chem. B} \textbf{2002}, \emph{106},
  11026--11035\relax
\mciteBstWouldAddEndPuncttrue
\mciteSetBstMidEndSepPunct{\mcitedefaultmidpunct}
{\mcitedefaultendpunct}{\mcitedefaultseppunct}\relax
\EndOfBibitem
\bibitem[Rizzo et~al.(2006)Rizzo, Aynechi, Case, and Kuntz]{Rizzo2006}
Rizzo,~R.; Aynechi,~T.; Case,~D.; Kuntz,~I. Estimation of absolute free
  energies of hydration using continuum methods: accuracy of partial, charge
  models and optimization of nonpolar contributions. \emph{J. Chem. Theor.
  Comput.} \textbf{2006}, \emph{2}, 128--139\relax
\mciteBstWouldAddEndPuncttrue
\mciteSetBstMidEndSepPunct{\mcitedefaultmidpunct}
{\mcitedefaultendpunct}{\mcitedefaultseppunct}\relax
\EndOfBibitem
\bibitem[Draper et~al.(2005)Draper, Grilley, and Soto]{Draper2005}
Draper,~D.; Grilley,~D.; Soto,~A. Ions and {RNA} folding. \emph{Annu. Rev.
  Biophys. Biomol. Struct.} \textbf{2005}, \emph{34}, 221--243\relax
\mciteBstWouldAddEndPuncttrue
\mciteSetBstMidEndSepPunct{\mcitedefaultmidpunct}
{\mcitedefaultendpunct}{\mcitedefaultseppunct}\relax
\EndOfBibitem
\bibitem[Chen et~al.(2009)Chen, Marucho, Baker, and Pappu]{Chen2009}
Chen,~A.; Marucho,~M.; Baker,~N.; Pappu,~R. Simulations of {RNA} interactions
  with monovalent ions. \emph{Methods Enzymol.} \textbf{2009}, \emph{469},
  411--432\relax
\mciteBstWouldAddEndPuncttrue
\mciteSetBstMidEndSepPunct{\mcitedefaultmidpunct}
{\mcitedefaultendpunct}{\mcitedefaultseppunct}\relax
\EndOfBibitem
\bibitem[Mackerell and Nilsson(2008)Mackerell, and
  Nilsson]{MackerellNilsson2008}
Mackerell,~A.; Nilsson,~L. Molecular Dynamics Simulations of Nucleic
  Acid-protein Complexes. \emph{Curr. Opin. Struct. Biol.} \textbf{2008},
  \emph{18}, 194--199\relax
\mciteBstWouldAddEndPuncttrue
\mciteSetBstMidEndSepPunct{\mcitedefaultmidpunct}
{\mcitedefaultendpunct}{\mcitedefaultseppunct}\relax
\EndOfBibitem
\bibitem[Shaw et~al.(2010)Shaw, Maragakis, Lindorff-Larsen, Piana, Dror,
  Eastwood, Bank, Jumper, Salmon, Shan, and Wriggers]{Shaw2010}
Shaw,~D.; Maragakis,~P.; Lindorff-Larsen,~K.; Piana,~S.; Dror,~R.;
  Eastwood,~M.; Bank,~J.; Jumper,~J.; Salmon,~J.; Shan,~Y.; Wriggers,~W.
  Atomic-Level Characterization of the Structural Dynamics of Proteins.
  \emph{Science} \textbf{2010}, \emph{330}, 341--346\relax
\mciteBstWouldAddEndPuncttrue
\mciteSetBstMidEndSepPunct{\mcitedefaultmidpunct}
{\mcitedefaultendpunct}{\mcitedefaultseppunct}\relax
\EndOfBibitem
\bibitem[Lindorff-Larsen et~al.(2011)Lindorff-Larsen, Piana, Dror, and
  Shaw]{Lindorff2011}
Lindorff-Larsen,~K.; Piana,~S.; Dror,~R.; Shaw,~D. How Fast-Folding Proteins
  Fold. \emph{Science} \textbf{2011}, \emph{334}, 517--520\relax
\mciteBstWouldAddEndPuncttrue
\mciteSetBstMidEndSepPunct{\mcitedefaultmidpunct}
{\mcitedefaultendpunct}{\mcitedefaultseppunct}\relax
\EndOfBibitem
\bibitem[Torrie and Valleau(1977)Torrie, and Valleau]{TorrieValleau1977}
Torrie,~G.; Valleau,~J. Nonphysical sampling distributions in {M}onte {C}arlo
  free-energy estimation: {U}mbrella sampling. \emph{J. Comput. Phys.}
  \textbf{1977}, \emph{23}, 187--199\relax
\mciteBstWouldAddEndPuncttrue
\mciteSetBstMidEndSepPunct{\mcitedefaultmidpunct}
{\mcitedefaultendpunct}{\mcitedefaultseppunct}\relax
\EndOfBibitem
\bibitem[Mezei(1987)]{Mezei1987}
Mezei,~M. Adaptive umbrella sampling: {S}elf-consistent determination of the
  non-{B}oltzmann bias. \emph{J. Comput. Phys.} \textbf{1987}, \emph{68},
  237--248\relax
\mciteBstWouldAddEndPuncttrue
\mciteSetBstMidEndSepPunct{\mcitedefaultmidpunct}
{\mcitedefaultendpunct}{\mcitedefaultseppunct}\relax
\EndOfBibitem
\bibitem[Grubm\"uller et~al.(1996)Grubm\"uller, Heymann, and
  Tavan]{Grubmuller1996}
Grubm\"uller,~H.; Heymann,~B.; Tavan,~P. Ligand binding: molecular mechanics
  calculation of the streptavidin-biotin rupture force. \emph{Science}
  \textbf{1996}, \emph{271}, 997--999\relax
\mciteBstWouldAddEndPuncttrue
\mciteSetBstMidEndSepPunct{\mcitedefaultmidpunct}
{\mcitedefaultendpunct}{\mcitedefaultseppunct}\relax
\EndOfBibitem
\bibitem[Darve and Pohorille(2001)Darve, and Pohorille]{DarvePohorille2001}
Darve,~E.; Pohorille,~A. Calculating free energies using average force.
  \emph{J. Chem. Phys.} \textbf{2001}, \emph{115}, 9169--9183\relax
\mciteBstWouldAddEndPuncttrue
\mciteSetBstMidEndSepPunct{\mcitedefaultmidpunct}
{\mcitedefaultendpunct}{\mcitedefaultseppunct}\relax
\EndOfBibitem
\bibitem[Laio and Parrinello(2002)Laio, and Parrinello]{LaioParrinello2002}
Laio,~A.; Parrinello,~M. Escaping free-energy minima. \emph{Proc. Natl. Acad.
  Sci. U.S.A.} \textbf{2002}, \emph{99}, 12562--12566\relax
\mciteBstWouldAddEndPuncttrue
\mciteSetBstMidEndSepPunct{\mcitedefaultmidpunct}
{\mcitedefaultendpunct}{\mcitedefaultseppunct}\relax
\EndOfBibitem
\bibitem[Rosso et~al.(2002)Rosso, Minary, Zhu, and Tuckerman]{Rosso2002}
Rosso,~L.; Minary,~P.; Zhu,~Z.; Tuckerman,~M. On the use of the adiabatic
  molecular dynamics technique in the calculation of free energy pro{fi}les.
  \emph{J. Chem. Phys.} \textbf{2002}, \emph{116}, 4389--4402\relax
\mciteBstWouldAddEndPuncttrue
\mciteSetBstMidEndSepPunct{\mcitedefaultmidpunct}
{\mcitedefaultendpunct}{\mcitedefaultseppunct}\relax
\EndOfBibitem
\bibitem[Marsili et~al.(2006)Marsili, Barducci, Chelli, Procacci, and
  Schettino]{Marsili2006}
Marsili,~S.; Barducci,~A.; Chelli,~R.; Procacci,~P.; Schettino,~V. Self-healing
  Umbrella Sampling: {A} Non-equilibrium Approach for Quantitative Free Energy
  Calculations. \emph{J. Phys. Chem. B} \textbf{2006}, \emph{110},
  14011--14013\relax
\mciteBstWouldAddEndPuncttrue
\mciteSetBstMidEndSepPunct{\mcitedefaultmidpunct}
{\mcitedefaultendpunct}{\mcitedefaultseppunct}\relax
\EndOfBibitem
\bibitem[Maragliano and Vanden-Eijnden(2006)Maragliano, and
  Vanden-Eijnden]{Maragliano2006}
Maragliano,~L.; Vanden-Eijnden,~E. A temperature accelerated method for
  sampling free energy and determining reaction pathways in rare events
  simulations. \emph{Chem. Phys. Lett.} \textbf{2006}, \emph{426},
  168--175\relax
\mciteBstWouldAddEndPuncttrue
\mciteSetBstMidEndSepPunct{\mcitedefaultmidpunct}
{\mcitedefaultendpunct}{\mcitedefaultseppunct}\relax
\EndOfBibitem
\bibitem[Sotomayor and Schulten(2007)Sotomayor, and
  Schulten]{SotomayorSchulten2007}
Sotomayor,~M.; Schulten,~K. Single-Molecule Experiments in Vitro and in Silico.
  \emph{Science} \textbf{2007}, \emph{316}, 1144--1148\relax
\mciteBstWouldAddEndPuncttrue
\mciteSetBstMidEndSepPunct{\mcitedefaultmidpunct}
{\mcitedefaultendpunct}{\mcitedefaultseppunct}\relax
\EndOfBibitem
\bibitem[Barducci et~al.(2008)Barducci, Bussi, and Parrinello]{Barducci2008}
Barducci,~A.; Bussi,~G.; Parrinello,~M. Well-Tempered Metadynamics: A Smoothly
  Converging and Tunable Free-Energy Method. \emph{Phys. Rev. Lett.}
  \textbf{2008}, \emph{100}, 020603--020606\relax
\mciteBstWouldAddEndPuncttrue
\mciteSetBstMidEndSepPunct{\mcitedefaultmidpunct}
{\mcitedefaultendpunct}{\mcitedefaultseppunct}\relax
\EndOfBibitem
\bibitem[Bartels and Karplus(1998)Bartels, and Karplus]{BartelsKarplus1998}
Bartels,~C.; Karplus,~M. Probability Distributions for Complex Systems:
  {A}daptive Umbrella Sampling of the Potential Energy. \emph{J. Phys. Chem. B}
  \textbf{1998}, \emph{102}, 865--880\relax
\mciteBstWouldAddEndPuncttrue
\mciteSetBstMidEndSepPunct{\mcitedefaultmidpunct}
{\mcitedefaultendpunct}{\mcitedefaultseppunct}\relax
\EndOfBibitem
\bibitem[Wang and Landau(2001)Wang, and Landau]{WangLandau2001}
Wang,~F.; Landau,~D. Efficient, Multiple-Range Random Walk Algorithm to
  Calculate the Density of States. \emph{Phys. Rev. Lett.} \textbf{2001},
  \emph{86}, 2050--2053\relax
\mciteBstWouldAddEndPuncttrue
\mciteSetBstMidEndSepPunct{\mcitedefaultmidpunct}
{\mcitedefaultendpunct}{\mcitedefaultseppunct}\relax
\EndOfBibitem
\bibitem[Micheletti et~al.(2004)Micheletti, Laio, and
  Parrinello]{Micheletti2004}
Micheletti,~C.; Laio,~A.; Parrinello,~M. Reconstructing the Density of States
  by History-Dependent Metadynamics. \emph{Phys. Rev. Lett.} \textbf{2004},
  \emph{92}, 170601--170604\relax
\mciteBstWouldAddEndPuncttrue
\mciteSetBstMidEndSepPunct{\mcitedefaultmidpunct}
{\mcitedefaultendpunct}{\mcitedefaultseppunct}\relax
\EndOfBibitem
\bibitem[Michel et~al.(2009)Michel, Laio, and Milet]{Michel2009}
Michel,~C.; Laio,~A.; Milet,~A. Tracing the Entropy along a Reactive Pathway:
  The Energy As a Generalized Reaction Coordinate. \emph{J. Chem. Theory
  Comput.} \textbf{2009}, \emph{5}, 2193--2196\relax
\mciteBstWouldAddEndPuncttrue
\mciteSetBstMidEndSepPunct{\mcitedefaultmidpunct}
{\mcitedefaultendpunct}{\mcitedefaultseppunct}\relax
\EndOfBibitem
\bibitem[Bonomi and Parrinello(2010)Bonomi, and
  Parrinello]{BonomiParrinello2010}
Bonomi,~M.; Parrinello,~M. Enhanced Sampling in the Well-Tempered Ensemble.
  \emph{Phys. Rev. Lett.} \textbf{2010}, \emph{104}, 190601--190604\relax
\mciteBstWouldAddEndPuncttrue
\mciteSetBstMidEndSepPunct{\mcitedefaultmidpunct}
{\mcitedefaultendpunct}{\mcitedefaultseppunct}\relax
\EndOfBibitem
\bibitem[Sugita and Okamoto(1999)Sugita, and Okamoto]{SugitaOkamoto1999}
Sugita,~Y.; Okamoto,~Y. Replica-exchange molecular dynamics method for protein
  folding. \emph{Chem. Phys. Lett.} \textbf{1999}, \emph{314}, 141--151\relax
\mciteBstWouldAddEndPuncttrue
\mciteSetBstMidEndSepPunct{\mcitedefaultmidpunct}
{\mcitedefaultendpunct}{\mcitedefaultseppunct}\relax
\EndOfBibitem
\bibitem[Marinari and Parisi(1992)Marinari, and Parisi]{MarinariParisi1992}
Marinari,~E.; Parisi,~G. Simulated Tempering: {A} New {M}onte {C}arlo Scheme.
  \emph{Europhys. Lett.} \textbf{1992}, \emph{19}, 451--458\relax
\mciteBstWouldAddEndPuncttrue
\mciteSetBstMidEndSepPunct{\mcitedefaultmidpunct}
{\mcitedefaultendpunct}{\mcitedefaultseppunct}\relax
\EndOfBibitem
\bibitem[Berg and Neuhaus(1992)Berg, and Neuhaus]{BergNeuhaus1992}
Berg,~B.; Neuhaus,~T. Multicanonical ensemble: A new approach to simulate
  first-order phase transitions. \emph{Phys. Rev. Lett.} \textbf{1992},
  \emph{68}, 9--12\relax
\mciteBstWouldAddEndPuncttrue
\mciteSetBstMidEndSepPunct{\mcitedefaultmidpunct}
{\mcitedefaultendpunct}{\mcitedefaultseppunct}\relax
\EndOfBibitem
\bibitem[Dingwall et~al.(1989)Dingwall, Ernberg, Gait, Green, Heaphy, Karn,
  Lowe, Singh, Skinner, and Valerio]{Dingwall1989}
Dingwall,~C.; Ernberg,~I.; Gait,~M.; Green,~S.; Heaphy,~S.; Karn,~J.; Lowe,~A.;
  Singh,~M.; Skinner,~M.; Valerio,~R. Human immunodeilciency virus {I} {T}at
  protein binds trans-activation-responsive region ({TAR}) {RNA} in vitro.
  \emph{Proc. Natl Acad. Sci. U.S.A.} \textbf{1989}, \emph{86},
  6925--6929\relax
\mciteBstWouldAddEndPuncttrue
\mciteSetBstMidEndSepPunct{\mcitedefaultmidpunct}
{\mcitedefaultendpunct}{\mcitedefaultseppunct}\relax
\EndOfBibitem
\bibitem[Dingwall et~al.(1990)Dingwall, Ernberg, Gait, Green, Heaphy, Karn,
  Lowe, Singh, and Skinner]{Dingwall1990}
Dingwall,~C.; Ernberg,~I.; Gait,~M.; Green,~S.; Heaphy,~S.; Karn,~J.; Lowe,~A.;
  Singh,~M.; Skinner,~M. {HIV}-1 {T}at Protein Stimulates Transcription by
  Binding to a {U}-rich Bulge in the Stem of the {TAR} {RNA} Structure.
  \emph{EMBO J.} \textbf{1990}, \emph{9}, 4145--4153\relax
\mciteBstWouldAddEndPuncttrue
\mciteSetBstMidEndSepPunct{\mcitedefaultmidpunct}
{\mcitedefaultendpunct}{\mcitedefaultseppunct}\relax
\EndOfBibitem
\bibitem[Weeks et~al.(1990)Weeks, Ampe, Schultz, Steitz, and
  Crothers]{Weeks1990}
Weeks,~K.; Ampe,~C.; Schultz,~S.; Steitz,~T.; Crothers,~D. Fragments of the
  {HIV}-1 {T}at Protein Specifically Bind {TAR} {RNA}. \emph{Science}
  \textbf{1990}, \emph{249}, 1281--1285\relax
\mciteBstWouldAddEndPuncttrue
\mciteSetBstMidEndSepPunct{\mcitedefaultmidpunct}
{\mcitedefaultendpunct}{\mcitedefaultseppunct}\relax
\EndOfBibitem
\bibitem[Calnan et~al.(1991)Calnan, Tidor, Biancalana, Hudson, and
  Frankel]{Calnan1991}
Calnan,~B.; Tidor,~B.; Biancalana,~S.; Hudson,~D.; Frankel,~A.
  Arginine-mediated {RNA} Recognition: the Arginine Fork. \emph{Science}
  \textbf{1991}, \emph{252}, 1167--1171\relax
\mciteBstWouldAddEndPuncttrue
\mciteSetBstMidEndSepPunct{\mcitedefaultmidpunct}
{\mcitedefaultendpunct}{\mcitedefaultseppunct}\relax
\EndOfBibitem
\bibitem[Delling et~al.(1992)Delling, Reid, Barnett, Ma, Climie, Sumner-Smith,
  and Sonenberg]{Delling1992}
Delling,~U.; Reid,~L.; Barnett,~R.; Ma,~M.; Climie,~S.; Sumner-Smith,~M.;
  Sonenberg,~N. Conserved nucleotides in the {TAR} {RNA} stem of human
  immunodeficiency virus type 1 are critical for {T}at binding and trans
  activationmodel for {TAR} {RNA} tertiary structure. \emph{J. Virol.}
  \textbf{1992}, \emph{66}, 3018--3025\relax
\mciteBstWouldAddEndPuncttrue
\mciteSetBstMidEndSepPunct{\mcitedefaultmidpunct}
{\mcitedefaultendpunct}{\mcitedefaultseppunct}\relax
\EndOfBibitem
\bibitem[Huthoff and Berkhout(2001)Huthoff, and Berkhout]{Huthoff2001}
Huthoff,~H.; Berkhout,~B. Mutations in the {TAR} hairpin affect the equilibrium
  between alternative conformations of the {HIV}-1 leader {RNA}. \emph{Nucleic
  Acids Res.} \textbf{2001}, \emph{29}, 2594--2600\relax
\mciteBstWouldAddEndPuncttrue
\mciteSetBstMidEndSepPunct{\mcitedefaultmidpunct}
{\mcitedefaultendpunct}{\mcitedefaultseppunct}\relax
\EndOfBibitem
\bibitem[Davidson et~al.(2009)Davidson, Leeper, Athanassiou, Patora-Komisarska,
  Karn, Robinson, and Varani]{Davidson2009}
Davidson,~A.; Leeper,~T.; Athanassiou,~Z.; Patora-Komisarska,~K.; Karn,~J.;
  Robinson,~J.; Varani,~G. Simultaneous Recognition of {HIV}-1 {TAR} {RNA}
  Bulge and Loop Sequences by Cyclic Peptide Mimics of {T}at Protein.
  \emph{Proc. Natl. Acad. Sci. U.S.A.} \textbf{2009}, \emph{106},
  11931--11936\relax
\mciteBstWouldAddEndPuncttrue
\mciteSetBstMidEndSepPunct{\mcitedefaultmidpunct}
{\mcitedefaultendpunct}{\mcitedefaultseppunct}\relax
\EndOfBibitem
\bibitem[Do et~al.(2012)Do, Ippoliti, Carloni, Varani, and Parrinello]{Do2012}
Do,~T.; Ippoliti,~E.; Carloni,~P.; Varani,~G.; Parrinello,~M. Counterion
  Redistribution upon Binding of a {T}at-Protein Mimic to {HIV}-1 {TAR} {RNA}.
  \emph{J. Chem. Theory Comput.} \textbf{2012}, \emph{8}, 688--694\relax
\mciteBstWouldAddEndPuncttrue
\mciteSetBstMidEndSepPunct{\mcitedefaultmidpunct}
{\mcitedefaultendpunct}{\mcitedefaultseppunct}\relax
\EndOfBibitem
\bibitem[Leach(2001)]{Leach2001}
Leach,~A. \emph{Molecular Modelling: Principles and Applications}, 2nd ed.;
  Prentice-Hall, 2001; p 604\relax
\mciteBstWouldAddEndPuncttrue
\mciteSetBstMidEndSepPunct{\mcitedefaultmidpunct}
{\mcitedefaultendpunct}{\mcitedefaultseppunct}\relax
\EndOfBibitem
\bibitem[Baker et~al.(2001)Baker, Sept, Joseph, Holst, and McCammon]{APBS2001}
Baker,~N.; Sept,~D.; Joseph,~S.; Holst,~M.; McCammon,~J. Electrostatics of
  nanosystems: {A}pplication to microtubules and the ribosome. \emph{Proc.
  Natl. Acad. Sci. U.S.A.} \textbf{2001}, \emph{98}, 10037--10041\relax
\mciteBstWouldAddEndPuncttrue
\mciteSetBstMidEndSepPunct{\mcitedefaultmidpunct}
{\mcitedefaultendpunct}{\mcitedefaultseppunct}\relax
\EndOfBibitem
\bibitem[Bonomi et~al.(2009)Bonomi, Branduardi, Bussi, Camilloni, Provasi,
  Raiteri, Donadio, Marinelli, Pietrucci, Broglia, and Parrinello]{PLUMED2009}
Bonomi,~M.; Branduardi,~D.; Bussi,~G.; Camilloni,~C.; Provasi,~D.; Raiteri,~P.;
  Donadio,~D.; Marinelli,~F.; Pietrucci,~F.; Broglia,~R.; Parrinello,~M.
  {PLUMED}: a portable plugin for free energy calculations with molecular
  dynamics. \emph{Comp. Phys. Comm.} \textbf{2009}, \emph{180},
  1961--1972\relax
\mciteBstWouldAddEndPuncttrue
\mciteSetBstMidEndSepPunct{\mcitedefaultmidpunct}
{\mcitedefaultendpunct}{\mcitedefaultseppunct}\relax
\EndOfBibitem
\bibitem[Minh and Adib(2008)Minh, and Adib]{MinhAdib2008}
Minh,~D.; Adib,~A. Optimized Free Energies from Bidirectional Single-Molecule
  Force Spectroscopy. \emph{Phys. Rev. Lett.} \textbf{2008}, \emph{100},
  180602--180606\relax
\mciteBstWouldAddEndPuncttrue
\mciteSetBstMidEndSepPunct{\mcitedefaultmidpunct}
{\mcitedefaultendpunct}{\mcitedefaultseppunct}\relax
\EndOfBibitem
\bibitem[Bennett(1976)]{Bennett1976}
Bennett,~C. Efficient estimation of free energy differences from {M}onte
  {C}arlo data. \emph{J. Comput. Phys.} \textbf{1976}, \emph{22},
  245--268\relax
\mciteBstWouldAddEndPuncttrue
\mciteSetBstMidEndSepPunct{\mcitedefaultmidpunct}
{\mcitedefaultendpunct}{\mcitedefaultseppunct}\relax
\EndOfBibitem
\bibitem[Bonomi et~al.(2009)Bonomi, Barducci, and Parrinello]{Bonomi2009}
Bonomi,~M.; Barducci,~A.; Parrinello,~M. Reconstructing the equilibrium
  {B}oltzmann distribution from well-tempered metadynamics. \emph{J. Comput.
  Chem.} \textbf{2009}, \emph{30}, 1615--1621\relax
\mciteBstWouldAddEndPuncttrue
\mciteSetBstMidEndSepPunct{\mcitedefaultmidpunct}
{\mcitedefaultendpunct}{\mcitedefaultseppunct}\relax
\EndOfBibitem
\bibitem[Colizzi and Bussi(2012)Colizzi, and Bussi]{ColizziBussi2012}
Colizzi,~F.; Bussi,~G. {RNA} unwinding from reweighted pulling simulations.
  \emph{J. Am. Chem. Soc.} \textbf{2012}, \emph{134}, 5173--5179\relax
\mciteBstWouldAddEndPuncttrue
\mciteSetBstMidEndSepPunct{\mcitedefaultmidpunct}
{\mcitedefaultendpunct}{\mcitedefaultseppunct}\relax
\EndOfBibitem
\bibitem[Kumar et~al.(1992)Kumar, Rosenberg, Bouzida, Swendsen, and
  Kollman]{Kumar1992}
Kumar,~S.; Rosenberg,~J.; Bouzida,~D.; Swendsen,~R.; Kollman,~P. The weighted
  histogram analysis method for free-energy calculations on biomolecules. {I}.
  {T}he method. \emph{J. Comput. Chem.} \textbf{1992}, \emph{13},
  1011--1021\relax
\mciteBstWouldAddEndPuncttrue
\mciteSetBstMidEndSepPunct{\mcitedefaultmidpunct}
{\mcitedefaultendpunct}{\mcitedefaultseppunct}\relax
\EndOfBibitem
\bibitem[Souaille and Roux(2001)Souaille, and Roux]{SouailleRoux2001}
Souaille,~M.; Roux,~B. Extension to the weighted histogram analysis method:
  combining umbrella sampling with free energy calculations. \emph{Comput.
  Phys. Commun.} \textbf{2001}, \emph{135}, 40--57\relax
\mciteBstWouldAddEndPuncttrue
\mciteSetBstMidEndSepPunct{\mcitedefaultmidpunct}
{\mcitedefaultendpunct}{\mcitedefaultseppunct}\relax
\EndOfBibitem
\bibitem[Hess et~al.(2008)Hess, Kutzner, van~der Spoel, and
  Lindahl]{GROMACS2008}
Hess,~B.; Kutzner,~C.; van~der Spoel,~D.; Lindahl,~E. {GROMACS} 4: {A}lgorithms
  for Highly Efficient, Load-Balanced, and Scalable Molecular Simulation.
  \emph{J. Chem. Theory Comp.} \textbf{2008}, \emph{4}, 435--447\relax
\mciteBstWouldAddEndPuncttrue
\mciteSetBstMidEndSepPunct{\mcitedefaultmidpunct}
{\mcitedefaultendpunct}{\mcitedefaultseppunct}\relax
\EndOfBibitem
\bibitem[Jorgensen et~al.(1983)Jorgensen, Chandrasekhar, Madura, Impey, and
  Klein]{Jorgensen1983}
Jorgensen,~W.; Chandrasekhar,~J.; Madura,~J.; Impey,~R.; Klein,~M. Comparison
  of simple potential functions for simulating liquid water. \emph{J. Chem.
  Phys.} \textbf{1983}, \emph{79}, 926--935\relax
\mciteBstWouldAddEndPuncttrue
\mciteSetBstMidEndSepPunct{\mcitedefaultmidpunct}
{\mcitedefaultendpunct}{\mcitedefaultseppunct}\relax
\EndOfBibitem
\bibitem[Lindorff-Larsen et~al.(2010)Lindorff-Larsen, Piana, Palmo, Maragakis,
  Klepeis, Dror, and Shaw]{Lindorff2010}
Lindorff-Larsen,~K.; Piana,~S.; Palmo,~K.; Maragakis,~P.; Klepeis,~J.;
  Dror,~R.; Shaw,~D. Improved side-chain torsion potentials for the {A}mber
  ff99{SB} protein force field. \emph{Proteins} \textbf{2010}, \emph{78},
  1950--1958\relax
\mciteBstWouldAddEndPuncttrue
\mciteSetBstMidEndSepPunct{\mcitedefaultmidpunct}
{\mcitedefaultendpunct}{\mcitedefaultseppunct}\relax
\EndOfBibitem
\bibitem[Perez et~al.(2007)Perez, Marchan, Svozil, Sponer, Cheatham, Laughton,
  and Orozco]{Perez2007}
Perez,~A.; Marchan,~I.; Svozil,~D.; Sponer,~J.; Cheatham,~T.; Laughton,~C.;
  Orozco,~M. Refinement of the {AMBER} Force Field for Nucleic Acids: Improving
  the Description of $\alpha/\gamma$ Conformers. \emph{Biophys. J.}
  \textbf{2007}, \emph{92}, 3817--3829\relax
\mciteBstWouldAddEndPuncttrue
\mciteSetBstMidEndSepPunct{\mcitedefaultmidpunct}
{\mcitedefaultendpunct}{\mcitedefaultseppunct}\relax
\EndOfBibitem
\bibitem[Cornell et~al.(1995)Cornell, Cieplak, Bayly, Gould, Merz, Ferguson,
  Spellmeyer, Fox, Caldwell, and Kollman]{Cornell1995}
Cornell,~W.; Cieplak,~P.; Bayly,~C.; Gould,~I.; Merz,~K.; Ferguson,~D.;
  Spellmeyer,~D.; Fox,~T.; Caldwell,~J.; Kollman,~P. A Second Generation Force
  Field for the Simulation of Proteins, Nucleic Acids, and Organic Molecules.
  \emph{J. Am. Chem. Soc.} \textbf{1995}, \emph{117}, 5179--5197\relax
\mciteBstWouldAddEndPuncttrue
\mciteSetBstMidEndSepPunct{\mcitedefaultmidpunct}
{\mcitedefaultendpunct}{\mcitedefaultseppunct}\relax
\EndOfBibitem
\bibitem[Cheatham et~al.(1999)Cheatham, Cieplak, and Kollman]{CheathamIII1999}
Cheatham,~T.~I.; Cieplak,~P.; Kollman,~P. A Modified Version of the {C}ornell
  \emph{et al.} Force Field with Improved Sugar Pucker Phases and Helical
  Repeat. \emph{J. Biomol. Struct. Dyn.} \textbf{1999}, \emph{16},
  845--862\relax
\mciteBstWouldAddEndPuncttrue
\mciteSetBstMidEndSepPunct{\mcitedefaultmidpunct}
{\mcitedefaultendpunct}{\mcitedefaultseppunct}\relax
\EndOfBibitem
\bibitem[Wang et~al.(2000)Wang, Cieplak, and Kollman]{Wang2000}
Wang,~J.; Cieplak,~P.; Kollman,~P. How well does a restrained electrostatic
  potential ({RESP}) model perform in calculating conformational energies of
  organic and biological molecules? \emph{J. Comput. Chem.} \textbf{2000},
  \emph{21}, 1049--1074\relax
\mciteBstWouldAddEndPuncttrue
\mciteSetBstMidEndSepPunct{\mcitedefaultmidpunct}
{\mcitedefaultendpunct}{\mcitedefaultseppunct}\relax
\EndOfBibitem
\bibitem[Hornak et~al.(2006)Hornak, Abel, Okur, Strockbine, Roitberg, and
  Simmerling]{Hornak2006}
Hornak,~V.; Abel,~R.; Okur,~A.; Strockbine,~B.; Roitberg,~A.; Simmerling,~C.
  Comparison of multiple {A}mber force fields and development of improved
  protein backbone parameters. \emph{Proteins} \textbf{2006}, \emph{65},
  712--725\relax
\mciteBstWouldAddEndPuncttrue
\mciteSetBstMidEndSepPunct{\mcitedefaultmidpunct}
{\mcitedefaultendpunct}{\mcitedefaultseppunct}\relax
\EndOfBibitem
\bibitem[Savelyev and Papoian(2007)Savelyev, and Papoian]{SavelyevPapoian2007}
Savelyev,~A.; Papoian,~G. Inter-{DNA} Electrostatics from Explicit Solvent
  Molecular Dynamics Simulations. \emph{J. Am. Chem. Soc.} \textbf{2007},
  \emph{129}, 6060--6061\relax
\mciteBstWouldAddEndPuncttrue
\mciteSetBstMidEndSepPunct{\mcitedefaultmidpunct}
{\mcitedefaultendpunct}{\mcitedefaultseppunct}\relax
\EndOfBibitem
\bibitem[Auffinger et~al.(2007)Auffinger, Cheatham, and Vaiana]{Auffinger2007}
Auffinger,~P.; Cheatham,~T.; Vaiana,~A. Spontaneous formation of {KCl}
  aggregates in biomolecular simulations: a force field issue? \emph{J. Chem.
  Theory Comput.} \textbf{2007}, \emph{3}, 1851--1859\relax
\mciteBstWouldAddEndPuncttrue
\mciteSetBstMidEndSepPunct{\mcitedefaultmidpunct}
{\mcitedefaultendpunct}{\mcitedefaultseppunct}\relax
\EndOfBibitem
\bibitem[Chen and Pappu(2007)Chen, and Pappu]{ChenPappu}
Chen,~A.; Pappu,~R. Parameters of monovalent ions in the {AMBER}-99 forcefield:
  Assessment of inaccuracies and proposed improvement. \emph{J. Phys. Chem. B}
  \textbf{2007}, \emph{111}, 11884--11887\relax
\mciteBstWouldAddEndPuncttrue
\mciteSetBstMidEndSepPunct{\mcitedefaultmidpunct}
{\mcitedefaultendpunct}{\mcitedefaultseppunct}\relax
\EndOfBibitem
\bibitem[Joung and Cheatham~III(2008)Joung, and
  Cheatham~III]{JoungCheatham2008}
Joung,~I.; Cheatham~III,~T. Determination of alkali and halide monovalent ion
  parameters for use in explicitly solvated biomolecular simulations. \emph{J.
  Phys. Chem. B} \textbf{2008}, \emph{112}, 9020--9041\relax
\mciteBstWouldAddEndPuncttrue
\mciteSetBstMidEndSepPunct{\mcitedefaultmidpunct}
{\mcitedefaultendpunct}{\mcitedefaultseppunct}\relax
\EndOfBibitem
\bibitem[Bussi et~al.(2007)Bussi, Donadio, and Parrinello]{Bussi2007}
Bussi,~G.; Donadio,~D.; Parrinello,~M. Canonical sampling through velocity
  rescaling. \emph{J. Chem. Phys.} \textbf{2007}, \emph{126},
  014101--014107\relax
\mciteBstWouldAddEndPuncttrue
\mciteSetBstMidEndSepPunct{\mcitedefaultmidpunct}
{\mcitedefaultendpunct}{\mcitedefaultseppunct}\relax
\EndOfBibitem
\bibitem[Parrinello and Rahman(1981)Parrinello, and
  Rahman]{ParrinelloRahman1981}
Parrinello,~M.; Rahman,~A. Polymorphic transitions in single crystals: {A} new
  molecular dynamics method. \emph{J. Appl. Phys.} \textbf{1981}, \emph{52},
  7182--7190\relax
\mciteBstWouldAddEndPuncttrue
\mciteSetBstMidEndSepPunct{\mcitedefaultmidpunct}
{\mcitedefaultendpunct}{\mcitedefaultseppunct}\relax
\EndOfBibitem
\bibitem[Jarzynski(1997)]{Jarzynski1997}
Jarzynski,~C. Nonequilibrium equality for free energy differences. \emph{Phys.
  Rev. Lett.} \textbf{1997}, \emph{78}, 2690--2693\relax
\mciteBstWouldAddEndPuncttrue
\mciteSetBstMidEndSepPunct{\mcitedefaultmidpunct}
{\mcitedefaultendpunct}{\mcitedefaultseppunct}\relax
\EndOfBibitem
\bibitem[Hummer and Szabo(2001)Hummer, and Szabo]{HummerSzabo2001}
Hummer,~G.; Szabo,~A. Free energy reconstruction from nonequilibrium
  single-molecule pulling experiments. \emph{Proc. Natl. Acad. Sci. U.S.A.}
  \textbf{2001}, \emph{98}, 3658--3661\relax
\mciteBstWouldAddEndPuncttrue
\mciteSetBstMidEndSepPunct{\mcitedefaultmidpunct}
{\mcitedefaultendpunct}{\mcitedefaultseppunct}\relax
\EndOfBibitem
\bibitem[Reynolds et~al.(1992)Reynolds, Essex, and Richards]{Reynolds1992}
Reynolds,~C.; Essex,~J.; Richards,~W. Atomic charges for variable molecular
  conformations. \emph{J. Am. Chem. Soc.} \textbf{1992}, \emph{114},
  9075--9079\relax
\mciteBstWouldAddEndPuncttrue
\mciteSetBstMidEndSepPunct{\mcitedefaultmidpunct}
{\mcitedefaultendpunct}{\mcitedefaultseppunct}\relax
\EndOfBibitem
\bibitem[Cieplak et~al.(1995)Cieplak, Cornell, Bayly, and Kollman]{Cieplak1995}
Cieplak,~P.; Cornell,~W.; Bayly,~C.; Kollman,~P. Application of the
  multimolecule and multiconformational {RESP} methodology to biopolymers:
  Charge derivation for {DNA}, {RNA}, and proteins. \emph{J. Comput. Chem.}
  \textbf{1995}, \emph{16}, 1357--1377\relax
\mciteBstWouldAddEndPuncttrue
\mciteSetBstMidEndSepPunct{\mcitedefaultmidpunct}
{\mcitedefaultendpunct}{\mcitedefaultseppunct}\relax
\EndOfBibitem
\bibitem[Orozco et~al.(2008)Orozco, Noy, and Perez]{Orozco2008}
Orozco,~M.; Noy,~A.; Perez,~A. Recent advances in the study of nucleic acid
  f{}lexibility by molecular dynamics. \emph{Curr. Opin. Struc. Biol.}
  \textbf{2008}, \emph{18}, 185--193\relax
\mciteBstWouldAddEndPuncttrue
\mciteSetBstMidEndSepPunct{\mcitedefaultmidpunct}
{\mcitedefaultendpunct}{\mcitedefaultseppunct}\relax
\EndOfBibitem
\bibitem[Banas et~al.(2010)Banas, Hollas, Zgarbova, Jurecka, Orozco, Cheatham,
  Sponer, and Otyepka]{Banas2010}
Banas,~P.; Hollas,~D.; Zgarbova,~M.; Jurecka,~P.; Orozco,~M.; Cheatham,~T.;
  Sponer,~J.; Otyepka,~M. Performance of Molecular Mechanics Force Fields for
  {RNA} Simulations: {S}tability of {UUCG} and {GNRA} Hairpins. \emph{J. Chem.
  Theory Comput.} \textbf{2010}, \emph{6}, 3836--3849\relax
\mciteBstWouldAddEndPuncttrue
\mciteSetBstMidEndSepPunct{\mcitedefaultmidpunct}
{\mcitedefaultendpunct}{\mcitedefaultseppunct}\relax
\EndOfBibitem
\bibitem[Yildirim et~al.(2011)Yildirim, Stern, Tubbs, Kennedy, and
  Turner]{Yildirim2011}
Yildirim,~I.; Stern,~H.; Tubbs,~J.; Kennedy,~S.; Turner,~D. Benchmarking
  {AMBER} force fields for {RNA}: comparisons to {NMR} spectra for
  single-stranded r({GACC}) are improved by revised $\chi$ torsions. \emph{J.
  Phys. Chem. B} \textbf{2011}, \emph{115}, 9261--9270\relax
\mciteBstWouldAddEndPuncttrue
\mciteSetBstMidEndSepPunct{\mcitedefaultmidpunct}
{\mcitedefaultendpunct}{\mcitedefaultseppunct}\relax
\EndOfBibitem
\bibitem[Denning et~al.(2011)Denning, Priyakumar, Nilsson, and
  Mackerell]{Denning2011}
Denning,~E.; Priyakumar,~U.; Nilsson,~L.; Mackerell,~A. Impact of 2'-hydroxyl
  Sampling on the Conformational Properties of {RNA}: Update of the {CHARMM}
  All-atom Additive Force Field for {RNA}. \emph{J. Comput. Chem.}
  \textbf{2011}, \emph{32}, 1929--1943\relax
\mciteBstWouldAddEndPuncttrue
\mciteSetBstMidEndSepPunct{\mcitedefaultmidpunct}
{\mcitedefaultendpunct}{\mcitedefaultseppunct}\relax
\EndOfBibitem
\bibitem[Besseova et~al.(2009)Besseova, Otyepka, Reblova, and
  Sponer]{Besseova2009}
Besseova,~I.; Otyepka,~M.; Reblova,~K.; Sponer,~J. Dependence of {A}-{RNA}
  simulations on the choice of the force field and salt strength. \emph{Phys.
  Chem. Chem. Phys.} \textbf{2009}, \emph{11}, 10701--10711\relax
\mciteBstWouldAddEndPuncttrue
\mciteSetBstMidEndSepPunct{\mcitedefaultmidpunct}
{\mcitedefaultendpunct}{\mcitedefaultseppunct}\relax
\EndOfBibitem
\bibitem[Noy et~al.(2009)Noy, Soteras, Luque, and Orozco]{Noy2009}
Noy,~A.; Soteras,~I.; Luque,~F.; Orozco,~M. The impact of monovalent ion force
  field model in nucleic acids simulations. \emph{Phys. Chem. Chem. Phys.}
  \textbf{2009}, \emph{11}, 10596--10607\relax
\mciteBstWouldAddEndPuncttrue
\mciteSetBstMidEndSepPunct{\mcitedefaultmidpunct}
{\mcitedefaultendpunct}{\mcitedefaultseppunct}\relax
\EndOfBibitem
\bibitem[McDowell et~al.(2007)McDowell, Spackova, Sponer, and
  Walter]{McDowell2007}
McDowell,~S.; Spackova,~N.; Sponer,~J.; Walter,~N. Molecular Dynamics
  Simulations of {RNA}: {A}n {\it In Silico} Single Molecule Approach.
  \emph{Biopolymers} \textbf{2007}, \emph{85}, 169--184\relax
\mciteBstWouldAddEndPuncttrue
\mciteSetBstMidEndSepPunct{\mcitedefaultmidpunct}
{\mcitedefaultendpunct}{\mcitedefaultseppunct}\relax
\EndOfBibitem
\bibitem[Not()]{Note-1}
Our modifications to the PLUMED software are available on request
\mciteBstWouldAddEndPuncttrue
\mciteSetBstMidEndSepPunct{\mcitedefaultmidpunct}
{\mcitedefaultendpunct}{\mcitedefaultseppunct}\relax
\EndOfBibitem
\bibitem[Bashford and Case(2000)Bashford, and Case]{Bashford2000}
Bashford,~D.; Case,~D. Generalized {B}orn models of macromolecular solvation
  effects. \emph{Annu. Rev. Phys. Chem.} \textbf{2000}, \emph{51},
  129--152\relax
\mciteBstWouldAddEndPuncttrue
\mciteSetBstMidEndSepPunct{\mcitedefaultmidpunct}
{\mcitedefaultendpunct}{\mcitedefaultseppunct}\relax
\EndOfBibitem
\end{mcitethebibliography}

\end{document}